\documentclass[usenatbib,usegraphicx]{mn2e}  

\usepackage{Times}
\usepackage{color}

\def\plum#1{\textcolor{black}{#1}}

\def\red#1{\textcolor{black}{#1}}
\def\red#1{\textcolor{black}{#1}}
\def\redd#1{\textcolor{black}{#1}}
\def\olive#1{{\color[named]{Black}#1}}

\def\msun{\thinspace M_\odot}
\def\abra#1#2{[{\rm {#1}} / {\rm{#2}}]}
\def\feoh{[{\rm Fe}/{\rm H}]}

\def\WT{Welch's $t$-test}
\def\STV{statistical $t$-value}
\def\mhyph{\  \mathchar`- \, }
 
\def\mabra#1#2{[{\rm {#1}} / {\rm{#2}}]_{\rm m}}
\def\avabra#1#2{[{\rm {#1}} / {\rm{#2}}]_{\rm ave}}

\def\dmean#1#2{\Delta [{\rm {#1}} / {\rm{#2}}]_{\rm m}}
\def\dave#1#2{\Delta [{\rm {#1}} / {\rm{#2}}]_{\rm ave}}
\def\zmax{Z_{\rm max}}
\def\UMP{{\sl \redd{ump}}}
\def\VMP{{\sl \redd{vmp}}}
\def\EMP{{\sl \redd{emp}}}
\def\MP{{\sl \redd{mp}}}
\def\lesssim{\la }

\title[The Stellar Abundances for Galactic Archaeology (SAGA) Database III.]{The Stellar Abundances for Galactic Archaeology (SAGA) Database III  --- Analysis of Enrichment Histories for Elements and Two Modes of Star Formation during the Early Evolution of Milky Way }

\author[Yamada et al.]{Shimako Yamada$^{1}$\thanks{E-mail:yamada@astro1.sci.hokudai.ac.jp }, 
Takuma Suda$^{2}$, 
Yutaka Komiya$^{2}$, 
Wako Aoki$^{2}$, 
\newauthor 
 Masayuki Y. Fujimoto$^{3,4}$, 
\\
$^{1}$Department of Cosmoscience, Hokkaido University, Kita 10 Nishi 8, Kita-ku, Sapporo 060-0810, Japan \\
$^{2}$National Astronomical Observatory of Japan, Osawa 2-21-1, Mitaka, Tokyo 181-0015, Japan \\
$^{3}$Nuclear reaction data center, Graduate School of Science, Hokkaido University, Kita 10 Nishi 8, Kita-ku, Sapporo 060-0810, Japan \\
$^{4}$Visiting researcher, Faculty of Engineering Hokkai-gakuen University, 4-1-40, Asahimachi, Toyohira-ku, Sapporo 062-8605, Japan
}

\begin{document}
\date{Accepted; Received 2013 April xx; in original form 2013 January 21}

\pagerange{\pageref{firstpage}--\pageref{lastpage}} \pubyear{2013}

\maketitle

\label{firstpage}

\begin{abstract} 

We study the enrichment histories for nine elements, C, four $\alpha$-elements of Mg, Si, Ca, and Ti, Sc, and three iron-peak elements of Co, Ni, and Zn, by using a large number of stellar data, collected by the Stellar Abundances for Galactic Archaeology (SAGA) database. 
   We find statistically significant changes, or breaks, of the mean abundance ratios to iron at three metallicities of $\feoh \simeq -1.8$, $-2.2$, and $-3.3$.  
   Across the first one, the mean abundance ratios decrease with the metallicity by similar extents for all the elements with the sufficient data.   
   Across the latter two, downward trends with the metallicity are also detected but for limited elements, C, Co, Zn, and possibly Sc, and for two of Co and Zn, respectively.  
   The breaks define four stellar populations with the different abundance patters which are dominant in each metallicity range divided by the breaks, Pop~IIa, IIb, IIc, and IId in order of increasing metallicity. 
   We also explore their spatial distributions with the spectroscopic distances to demonstrate that Pops~IIa and IIb spread over the Galactic halo while Pops~IIc and IId are observed near the Galactic plane.  
   In particular, Pop~IIc stars emerge around $\feoh \simeq -2.6$ and coexist with Pop~IIb stars, segregated by the spatial distributions.   
   Our results reveal two distinct modes of star formation during the early stages of Galaxy formation, which are associated with the variations of IMF and the spatial distribution of remnant low-mass stars. 
   For the two lower-metallicity populations, the enhancements of Zn and Co indicate a high-mass and top-heavy IMF together with the statistics on the carbon-enhanced stars.   
   For the two higher-metallicity populations, on the other hand, the difference in the abundance patterns is attributable to the delayed contribution of SNe Ia, indicative of a low-mass IMF and the specific star formation rate comparable to the present-day Galactic disk.  
   We discuss the relevance to the kinematically resolved structures of the Galactic halo and the possible sites of these populations within the framework of hierarchical structure formation scenario.  
\end{abstract}


\section{Introduction}\label{sec:intro}
\renewcommand\thefootnote{\arabic{footnote})}

During past decades, a large number of extremely metal-poor (EMP) stars with the metallicity well below the lowest metallicity of globular clusters ($\feoh \lesssim -2.5$) have been discovered in the Galactic halo thanks to large-scaled HK and HES surveys \citep[][]{beers05}. 
   These stars should have been born out of gas polluted by the ejecta of the first and subsequent supernovae and survive to date because of low-masses ($M \la 0.8 \msun$).  
   Their surface abundances, unveiled by high-dispersion spectroscopic observations, may carry imprints of the nucleosynthesis and metal-enrichment process during the formation of the Galactic halo and serve as a probe into the structure formation in the early Universe.  

The Stellar Abundances for Galactic Archaeology (SAGA) database (http://saga.sci.hokudai.ac.jp/) assembles the abundance data and other stellar parameters, aiming at covering all the papers dealing with stars of $\feoh \le -2.5$, published since 2000 and some traced back to 1994 \citep[][referred to as Papers~I and II, respectively, in the following]{suda08,suda11}. 
   It enables to explore the attributes of EMP and metal-poor stars through the statistical analyses of a large sample of stars, observed to date. 
   In Paper~II, we discuss the overall characteristics of abundance peculiarities of EMP stars related to the internal and external elemental pollution including the carbon enhancement. 
   In this paper, we investigate detailed histories of chemical enrichment for elements as many as possible by using the abundances and other stellar parameters, to reveal the nucleosynthetic processes and sites in the early stage of the Galaxy formation. 

The enrichment histories of elements are connected to the star formation rate (SFR) and the initial mass function (IMF), coupled with the mass and metallicity dependences of the lifetimes and nucleosynthetic yields of supernovae (SNe). 
   The SFR may influence the element enrichment through the difference in the lifetimes of progenitors to SNe.  
   The core-collapsed SNe (CCSNe) enrich the interstellar medium with most of metals as well as iron in timescales of $\sim 10$ Myr or so, while Type Ia SNe dominantly eject iron in much longer timescales typically of $\sim 1$ Gyr.
   Accordingly, Type Ia SNe lag behind CCSNe and may bring about a decrease in the abundance ratio of metals to iron \citep[e.g.,][]{matteucci90,matteucci03,tsujimoto95}.   
   It is reported that stars exhibit decrease in the abundance ratios of $\alpha$-elements to iron around $\feoh \ga -2$ for dwarf spheroidal (dSph) galaxies \citep[e.g., see][]{tolstoy09,kirby11}, which is interpreted to mean that these systems achieve the metal-enrichments after Type Ia SNe have contributed to the chemical evolution. 
   The time lag of Type Ia SNe behind CCSNe may introduce the timescale to the metal-enrichment histories, as well known.

As for the IMF of EMP stars, constraint is discussed from the statistics of carbon enhanced stars in the Galactic halo \citep{lucatello05,komiya07}, based on the binary scenario for their origin \citep[][also see Suda et al 2013]{suda04}. 
   \citet{komiya07} deduce high-mass IMF with typical mass, $ M_{\rm md} \simeq 10 \msun$, 
   applying the theoretical results on the elemental mixing and nucleosynthesis during helium flashes, peculiar to EMP stars \citep{fujimoto90,hollowell90,fujimoto00,suda10}. 
   Furthermore, the same conclusion is derived from the scarcity of EMP stars, detected by HK and HES surveys \citep{komiya09a}. 
   If the high-mass IMF is the case for EMP stars, there should be transition to low-mass IMF such as observed for stars of younger populations during the Galaxy halo formation (see Paper II). 
   The IMF change should have left imprints of nucleosynthetic signatures on the surface abundances that stem from the dependence of SN yields on the progenitor mass.  

The chemical enrichment of Galactic halo has been studied by many authors.   
   As for $\alpha$-elements, a trend of the abundance ratio to iron, $\abra{\alpha}{Fe}$, decreasing with the metallicity is asserted \citep[e.g.,][]{stephens02, gratton03, zhang09}, which is claimed as indicative that the IMF is not invariant in time \citep{stephens02}. 
   Recently, \citet{ishigaki12} discuss a decrease of $\abra{\alpha}{Fe}$ between $\feoh \simeq -2.5$ and $-2$, while others deem it nearly constant \citep[e.g.,][]{nissen94,clementini99,cayrel04}.  
   \citet{nissen10} detect two distinct groups with different $\abra{\alpha}{Fe}$ among the halo stars of $\feoh \ga -1.5$, whereas \citet{ruchti10} argue the spread of abundances only for Mg. 
   In addition, larger mean of $\abra{\alpha}{Fe}$ is reported for a bulk of halo stars, and also, for the thick disk stars \citep[e.g.,][]{edvardsson93,prochaska00,mashonkina00,mashonkina01,reddy06,bensby03,bensby05,brewer06,ruchti10,ruchti11}, which is interpreted as the consequence of CCSNe preceding Type Ia SNe.  
   Among the iron-peak elements, Zn and Co are known to exhibit intriguing behaviors, different from  other elements. 
   A trend of their abundances increasing relative to iron monotonically for lower metallicity is asserted \citep{ryan96,mcwilliam95,primas00,johnson02,cayrel04,takeda05,saito09,bergemann10,honda11}.
   On the other hand, discontinuous change in $\abra{Zn}{Fe}$ around $\feoh \simeq -2$ is pointed out \citep{nissen04,nissen07}. 

The understanding of element enrichments is yet to be properly established, in particular with relation to SFR and IMF.  
   More extensive and systematic studies are necessary to draw a consistent picture on the chemical enrichment histories.  
   The previous studies are prone to suffer from small number statistics, which prevents from deriving reliable results.  
   The SAGA database, on the other hand, possesses larger sample data by more than an order of magnitude and we may rely on statistics to analyze the characteristics of stellar abundances.   
\red{
As assembled from various papers, the data are liable to systematic errors and/or selection biases among different authors, so that we have to proper attention to their possible effects in discussing the characteristics of collected abundances.  
   In this paper, we focus on the differential variations between the sample stars in the adjacent narrow metallicity ranges and, if necessary, may restrict the comparisons to stars of similar properties in an attempt to minimize the influences.  
}
   Considering possible non-linear nature of enrichments, we apply the \WT\ to inspect the difference in the mean abundance ratios to iron, which is also suitable to deal with the combined data from many works.  
   Furthermore, we take into account the spatial distributions of sample stars, provided by the SAGA database, and explore the structural characteristics of the enrichment processes.   
   Our main interest is to gain insights into SFR, IMF and SN nucleosynthesis during the formation and early evolution of the Galaxy by tracing the variations in the element enrichments, embedded in the assembly of observed data.

The paper is organized as follows. 
   In \S 2, we describe the selection of target elements among the sample data in the SAGA database and present the results of analysis of the enrichment histories by applying the \WT\ to selected elements. 
   In \S 3, the global features of element enrichment are summarized to formulate four stellar populations with different abundance patterns that constitute the sample stars.   
   In \S 4, we investigate the spatial distributions of sample stars and the structural characteristics of these populations.  
   The results are discussed in \S 5 in relation to the origins and transitions of four populations and the implications to the formation processes of the Galaxy. 
   Conclusions follow.

\section{Analysis of enrichment histories}\label{sec:analysis}



\begin{table*}
 \begin{center}
 \begin{minipage}{140mm}
    \caption{The numbers of sample stars with the abundance data of elements for $\feoh < -1$, registered in the SAGA database (October 31 2010 version)} 
\label{table:elements}
\vskip 5pt
\begin{tabular}{|c|c|c|c|c|c|c|c|}
\hline
  &  &  \multicolumn{6}{c}{high-resolution spectroscopic data by resolution, $R > 20,000$. } \\ 
  \cline{3-8} 
Elements & All data & & \multicolumn{5}{c}{Selected by carbon abundances and with outliers removed 
\footnote{(4-8th columns) Both carbon-enhanced and carbon-depleted stars (see text for their definition) are removed from the sample of Li, C, N, Na, and Al, and carbon-enhanced stars from the sample of $\alpha$-elements (O - Ti), Sc and neutron capture elements (Sr - Eu), denoted by $^{\dagger}$ and $^{*}$, respectively. 
   In addition, excluded are the outliers with the enhancements that deviate from the mean by more than $3 \sigma$ (see Appendix~\ref{app:outliers} for detail) and the blue metal-poor stars by \citet{preston00}} } \\  
\cline{4-8} 
   & & &  & RGB  &  MS  & $|Z| \ge 500$ pc & $ < 500$ pc   
\footnote{RGB and MS denote giants and dwarfs divided by the surface gravity of $\log g (\hbox{ dyn g}^{-1}) = 3.5$, and the last two columns are sorted by the height, $|Z|$, from the Galactic plane, where some stars are doubly counted because of the scatters of data.} \\
    \hline 
Li & 285   & 247    &  111   $^{\dagger}$  & 54  &  66  & - & - \\
C & 638    &  343  &   204 $^{\dagger}$  &  118  & 91  & 95 & 121  \\ 
N  & 147   & 134   &  52  $^{\dagger}$  & 43  & 12  & - & -  \\
O & 257    & 253  &   178,264,67 $^{*,}$
\footnote{The abundance data by [OI], OI triplet and OH lines, respectively.}   &  -  & -& - & - \\
Na & 354   & 353   &  113 $^{\dagger}$  & 89  & 61 & - & - \\
Mg & 766  &  488  &   354 $^{*}$  &  174   &   196 & -& 185 \\
Al & 487    & 262 &    117 $^{\dagger}$  & 85  & 41  &-&-  \\
Si & 331    &  331 &   280 (193) $^{*, d}$    &   136   &    70  & - & 148 \\
Ca & 773 & 481 &    347 $^{*}$  &   178  &   193 & - &     184   \\  
Sc & 499 & 264  & 193 $^{*}$  &   130  & - & - &    - \\    
Ti II & 444 & 387 &   258 (212) $^{*, d }$    & 124   &  123  &- &-  \\
V & 199 & 153 &     138 &  109 & -  & - & -    \\
Cr I (II) $^{ d }$ & 663 (116) & 404 (116) & 353 (114) & -  & -  & -  & -\\
Mn I (II) $^{ d}$ & 481(42) &  256 (42) &   205(41) & -  & -    & -  & - \\
Fe & 1035  &  644 &   -  &  -  & -  & -  & - \\
Co & 399  & 186  &    179  &   107   &  - & 163 & -  \\
Ni & 570 &337 &     333  &   171   & 181 &  176 &  173 \\
Zn & 237 & 200   & 179&    123(60)
\footnote{Figures in parenthesis represent the numbers of stars with the abundance data of $T_{\rm eff} \le 5800$ K for Si, with the dwarfs of $\feoh < -2.3$ excluded for Ti II, with the Cr II and Mn II abundances, and with Zn abundances studied by \citet{saito09}, \citet{nissen07}, \citet{johnson02}, and \citet{mishenina02}, respectively.} & 66(49)$ ^{ d}$    & 98 &  100   \\
Sr &600  &314   &   265 $^{*}$   & 110  & 169  & -  & - \\
Y & 384  & 232 &   200 $^{*}$   &  128  & 84  & -  & -\\
Zr &   209 & 157 &  140 $^{*}$  & 112   & 38  & -  & -\\
Ba & 714   & 432  &  365$^{*}$   &  201   & 202   & -  & -\\
La & 205   & 172 &  122  $^{*}$   &  113   &33  & -  & -\\
Nd  &  148  &  112 &   85$^{*}$   & 81   & 9  & -  & -\\ 
Eu  &  459  &  389 &  165 $^{*}$   & 129  & 41   & -  & - \\ 
\hline
\end{tabular}                 
\end{minipage}
\end{center}

\end{table*}


In this section, we study the enrichment histories that each element may have by utilizing the SAGA database (Oct. 31, 2010 version). 
   It resisters 1386 Galactic halo stars in total, taken from 158 papers which include at least a star of $\feoh \le -2.5$, and contains 23,775 independent reports on the abundances if the data for the same objects but from different papers are counted separately.  

It is known that the enrichment histories differ among elements \citep[e.g.,][]{cayrel04} and may be classified according to the variations of their abundances relative to iron.  
   In Paper II, we show that the elements are sorted into four groups through the linear fit to the relationship between the abundances of element and iron among the sample stars of $\feoh < -1$;  
   three groups with the tendency of the abundance ratio to iron (i) increasing, (ii) remaining nearly constant, and (iii) decreasing toward lower metallicity, respectively, and (iv) another group of neutron-capture elements heavier than iron (see Fig.~15 and Table~2 in Paper II). 
   The former three groups include (i) carbon, oxygen, two iron-peak elements (Co and Zn), (ii) $\alpha$-elements (Mg, Si, Ca, Ti), two odd atomic number elements (Sc and V) in between the $\alpha$- and iron-peak elements, an iron-peak element (Ni) and (iii) other iron-peak elements (Cr, Mn and Cu), respectively.   
   In the following, we call the abundance ratio to iron, $\abra{X}{Fe}$, "enhancement" for convenience' sake and distinguish it from the abundance ratio to hydrogen, $\abra{X}{H}$.  

In this paper, we study the variations of the mean enhancements with the metallicity by applying the \WT\ to examine the statistical significance of the variations between the sample stars in two adjacent metallicity ranges.  
\red{
   We work on the LTE abundances and try not to consider the NLTE and 3D corrections in order to use the homogenized data;  
   the latter effects may have the dependences on metallicity and other parameters \citep[e.g., see][]{asplund05}, and yet, we can reduce their influences by restricting the comparison to the sample stars within narrow metallicity ranges and of similar properties.  
}
   We also consider the resolution of spectroscopic observations since the abundance data by low-resolution spectroscopy are subject to larger errors.   

We select the objects of analysis as follows.  
   We exclude the stars whose surface abundances may have suffered modifications after birth through the internal mixing and nuclear burning and/or through the external pollution such as wind accretion from primary AGB stars in binary systems \citep[e.g.,][Paper~II]{suda04,komiya07}. 
   In particular, we remove the carbon-enhanced stars from the sample stars for the elements that may be affected by the nucleosynthesis in AGB stars.  
   Also excluded are the outliers with the abundances that deviate by more than 3 $\sigma$ from the mean enhancements among the sample stars in the same metallicity range.  

The result of sample selections are summarized in Table~\ref{table:elements} for the elements with more than 100 abundance data by high-resolution spectroscopy of $R > 20,000$. 
   Among them, O, V, Cr, and Mn are hard to draw any statistically significant consequences because of the discrepancies in the abundance determination from different lines, as discussed in Appendix~\ref{app:unanalyzed}. 
   We will not deal with the odd atomic number elements, Li, Na, and Al, either since their synthetic processes are thought to be subject to 'extra' mixing other than the thermal convection in the interior of stars.   
   The discussion of neutron capture elements is deferred to a forthcoming paper since they owe the enrichment not only to r-process nucleosynthesis during supernovae and other explosive events but also to s-process nucleosynthesis in AGB stars.  
   Consequently, we perform the detailed analysis for 9 elements, C, four $\alpha$-elements of Mg, Si, Ca, and Ti, an odd element, Sc, in between the $\alpha$- and iron-peak elements, three iron-peak elements of Co, Ni, and Zn.   

\begin{figure*}
\begin{center}
\includegraphics[width=200mm,height=210mm,bb=90 41 750 763, clip]{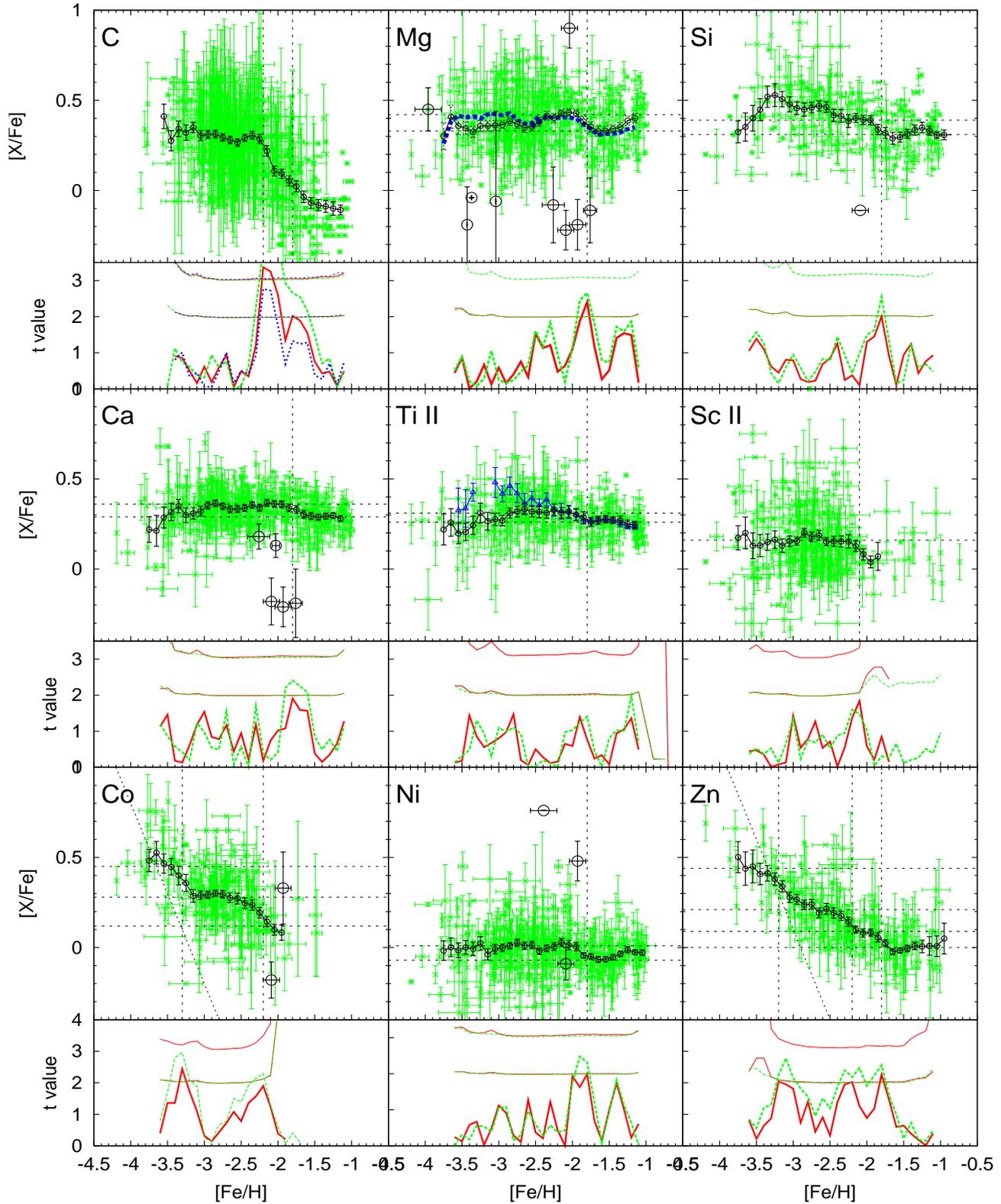}
\end{center}
\caption{ 
   The enrichment histories, imprinted among the sample stars of $\feoh<-1$, registered in the SAGA database, for selected nine elements.  
   Upper section of each panel shows the distributions of enhancements, $\abra{X}{Fe}$ against the metallicity, $\abra{Fe}{H}$, and also the mean enhancements for the sample stars in the bin width of 0.3 dex in $\feoh$ against the central metallicity of bins which include 5 and more sample stars. 
  Lower section shows the statistical $t$-values of \WT\ for the sample stars in the two adjacent metallicity bins of 0.3 dex or 0.4 dex width on either side (red and green lines, respectively), together with the critical $t$-values corresponding to the statistical significance levels of $2\sigma$ and $3\sigma$ (lower and upper lines that run nearly horizontally, respectively). 
   The result of \WT\ with 0.2 dex bin width is shown for carbon. 
   Large open circles denote outliers excluded from the analysis, which have the abundances different from the mean enhancement in the same metallicity range by more than $3 \sigma$ at least for some elements (see  Appendix~\ref{app:outliers} for details). 
}  
\label{fig:allfe}
\end{figure*}

The enrichment histories and the result of \WT\ are shown against the metallicity in Figure~\ref{fig:allfe} for all the elements selected.   
   Upper section of each panel presents the variation of mean enhancement for the sample stars in the metallicity bin of 0.3 dex width in $\feoh$ together with the enhancements of individual sample stars.  
   It exhibits overall tendencies of mean enhancements increasing toward lower metallicity or remaining nearly flat, as stated above. 
   \red{In addition, there are undulations superposed, whether real features inherent in the element enrichment or artifacts originating from the observation errors, especially the selection bias and/or systematic errors among the data taken from different authors. } 
   
Lower section presents the {\STV}s of the \WT\ for the sample stars in two metallicity sections on either side with the bin widths of 0.3 and 0.4 dex along with the critical $t$-values, $t_{2\sigma}$ and $t_{3\sigma}$, corresponding to the statistical significance of $2 \sigma$ and $3 \sigma$ levels.    
   It reveals that rather steep variations, or breaks, in the mean enhancement occur across three metallicities of $\feoh \simeq -1.8$, -2.2, and $-3.3 \mhyph -3.2$ with the statistical significance of $2\sigma$ and even $3 \sigma$ levels.  
   These breaks are found at the same metallicity within the errors for multiple elements, and in particular the first one for all the elements with the sufficient abundance data in the relevant metallicity ranges.  
   Between or outside the breaks, the sample stars are regarded as having the same, or slowly varying, mean enhancements within the error bars.  
   In the following, we refer to four metallicity divisions, separated by these three breaks, as ''metal-poor'' (\MP), ''very metal-poor'' (\VMP), ''extremely metal-poor'' (\EMP) and ''ultra metal-poor'' (\UMP) divisions in order of decreasing metallicity. 
   In this paper, we use lower cases for acronyms since the metallicity ranges slightly differ from those proposed by \citet{beers05}.  

We below describe the characteristics of enrichment histories and the occurrence of the breaks for three typical elements, carbon, magnesium of $\alpha$-elements, and zinc among the iron-peak elements.  
   The other elements are discussed in Appendix~\ref{app:analyzed}.  


\subsection{Carbon}\label{subsec:carbon}

It is known that a considerable fraction of EMP stars display the enhancement of carbon, affected by the mass transfer and/or the wind accretion from AGB stars in binary systems \citep[][]{rossi99,fujimoto00,suda04,lucatello05,komiya07}.  
   In addition, some of EMP giants exhibit carbon depletion with nitrogen enhancement \citep[mixed stars,][]{spite05}, which is indicative of the internal mixing of matter processed via CN cycles along with the fact that $\abra{C}{Fe}$ decreases as stars ascend the RGB \citep[e.g.,][]{cohen06,lai08}.  
   These peculiar stars are also observed among stars of heigher metallicity.  
   Here we deal only with the stars of ''normal'' carbon abundances, which are thought to be free from the modifications of surface abundances after birth. 
   As the criterion of carbon enhanced stars, we adopt $\abra{C}{Fe} \ge 0.7$ for the stars of $\feoh \le -2$ and slightly smaller $\abra{C}{Fe} \ge 0.5$ for $\feoh > -2$ according to \citet{aoki07} and Paper~II (see Fig.~2 of Paper~II).  
   The criterion of carbon depletion is taken to be $\abra{C}{Fe} \le -0.1$ and $-0.4$ for $\feoh \le -2$ and $> -2$, respectively (see Fig.~19 of Paper~II).  

The enrichment history for 411 carbon-normal stars, registered in the SAGA database in the range of $\feoh<-1$ is shown in Fig.~\ref{fig:allfe} (top-left panel).  
   The mean enhancement remains nearly flat with $\mabra{C}{Fe} \simeq 0.3$ for $\feoh \la -2.3$.
   Then, it starts to decline abruptly for higher metallicity and keeps decreasing at varying rates until gradually settling in nearly flat level with $\mabra{C}{Fe} \simeq - 0.1$ for $\feoh \ga -1.6$.  

The \WT\ shows that the {\STV}s exceed $t_{2 \sigma}$, or even $t_{3\sigma}$, over a rather wide range of metallicity around $\feoh \simeq -2$ both for the bin widths of 0.3 and 0.4 dex.  
   The results indicate that the null hypothesis that the sample stars have the same mean enhancement in two adjacent metallicity bins is rejected, by statistical significance levels more than $2 \sigma$ and $3 \sigma$, i.e., there should be variations in $\mabra{C}{Fe}$ across these metallicity ranges.  
   In generally, the \STV\ tends to be larger for larger bin width since the sample variance reduces with the number of sample stars.  
   In actuality, for the 0.4 dex bin width, it displays a broader peak of $t > 3 \sigma$ over $\feoh = -2.2 \mhyph -1.9$ and a shoulder of $t > t_{\rm 2 \sigma}$ extends up to $\feoh = -1.6$.  
   For the 0.3 dex bin width, on the other hand, it develops a tendency that the peak and shoulder are decomposed into two peaks, centered at $\feoh \simeq -2.2$ and $-1.8$, respectively. 
   For still smaller the 0.2 bin width, the \WT\ keeps the same tendency with the notch in the middle.   
   This implies a change in the decline rate, or an intervening constant part, of mean enhancement in between these metallicities.  
   Outside this metallicity range, the \STV\ sharply drops and remains very small without an apparent dependence on the bin width. 
   That is, the \WT\ fails to reject the null hypothesis, indicative that $\mabra{C}{Fe}$ between the adjacent metallicity bins remains same within errors over the corresponding metallicity range.  
   We perform the same analysis for 204 carbon-normal stars with the abundances, measured by high dispersion spectroscopy of resolution $R>20,000$ to find basically the same results.  


\begin{figure}
\begin{center}
\includegraphics[width=85mm,height=100mm,bb=0 20 400 500, clip]{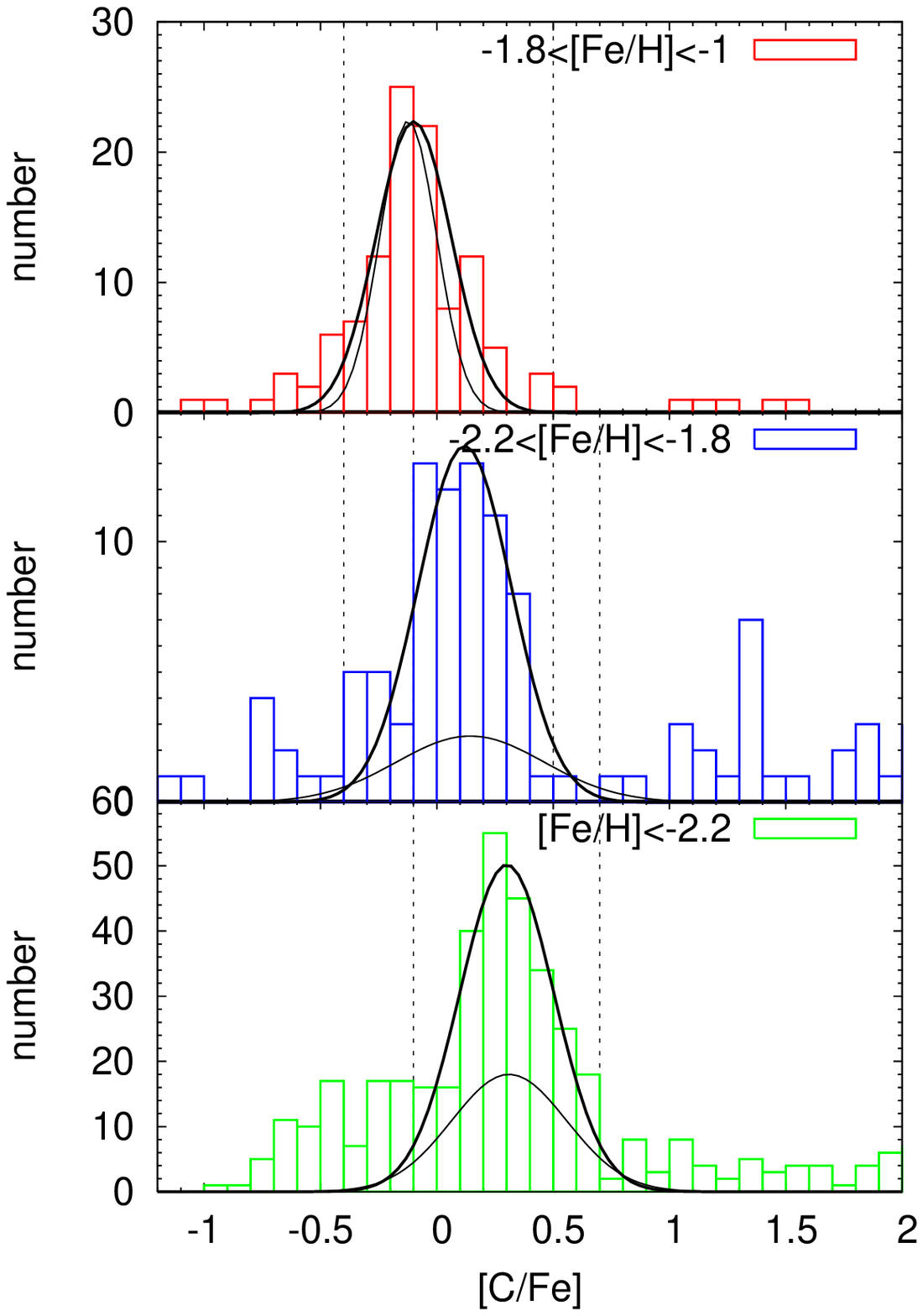}
\end{center}
\caption{ 
   Histograms of the carbon enhancement, $\abra{C}{Fe}$, for the sample stars in the metallicity ranges, divided by the breaks of $\feoh = -1.8$ and $-2.2$. 
   Dotted lines denote the borders of the carbon-normal stars with the carbon-enhanced or depleted stars (for their definitions, see the text).  
   Thick and thin lines show the Gaussian fitting to the distributions of carbon enhancement for all the carbon-normal stars and for those with the abundance data by high-resolution spectroscopy of resolution $R > 20,000$. 
\label{fig:hist-cfe} }
\end{figure}

The above result suggests that precipitous changes, or breaks, of the mean enhancement of carbon occur across the metallicity, centered at $\feoh = -2.2$ and $-1.8$, respectively.  
   Figure~\ref{fig:hist-cfe} compares the distributions of $\abra{C}{Fe}$ in three metallicity ranges, divided by these breaks.  
   Here we include both the carbon-enhanced and deficient stars, and yet, the carbon-normal stars stand out clearly against them, and are well fitted by Gaussian curves. 
   There are obvious shifts of the distributions toward smaller $\abra{C}{Fe}$ as the metallicity increases while the scatters are kept nearly similar (see Table~\ref{table:sigma} below).   
   Accordingly we can consider that the sample stars in these three divisions possess separate distributions of carbon enhancement, and in particular, that the distribution in the middle range is neither a superposition nor continuous change between those of two metallicity ranges on both sides.


\begin{figure}
\begin{center}
\includegraphics[width=95mm,height=50mm,bb=75 45 630 307, clip]{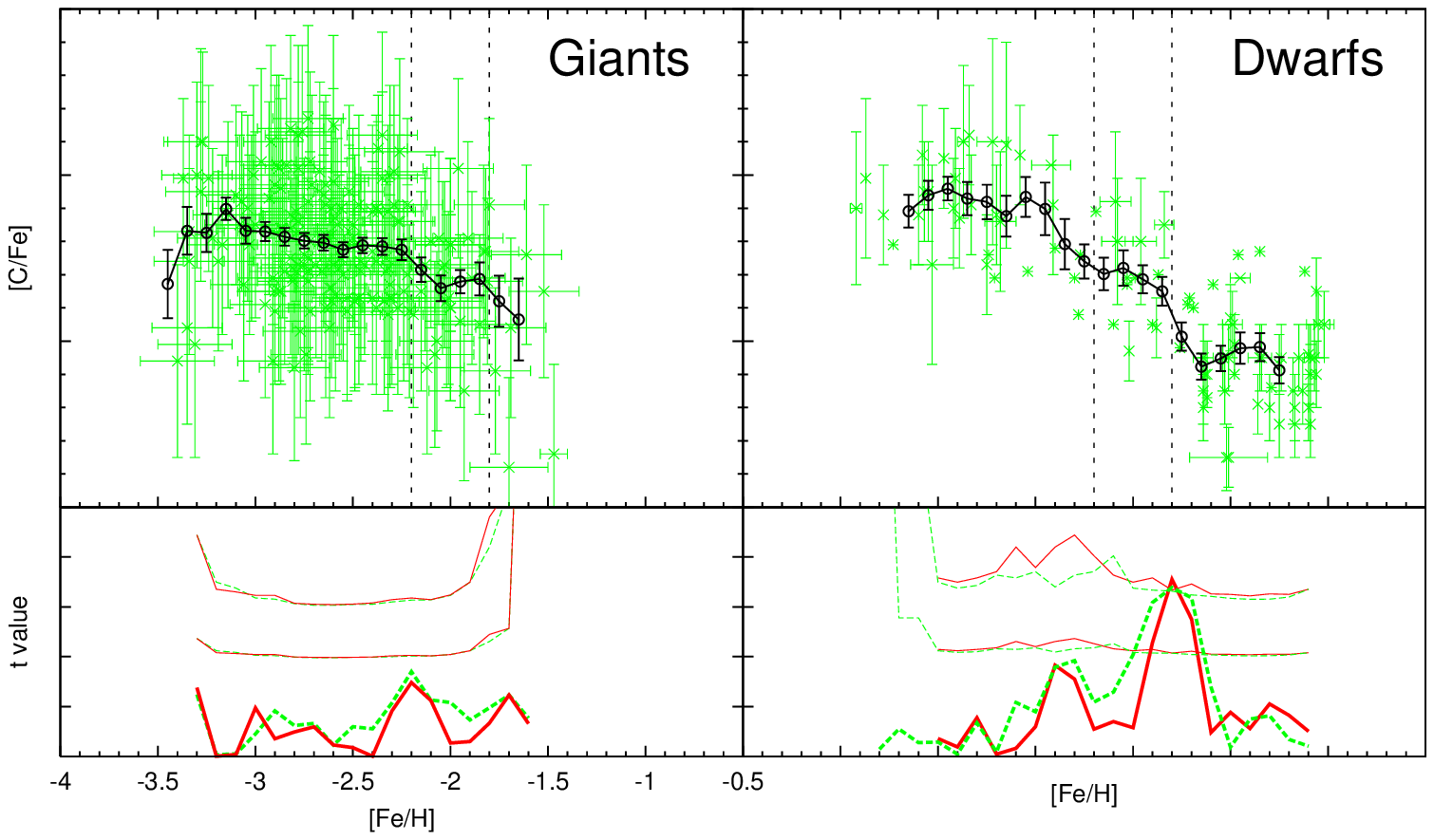}
\end{center}
\caption{The same as top-left panel of Fig~\ref{fig:allfe} but for the giants with the abundance data by \citet{barklem05} (left panel) and for the dwarfs of $\log g > 3.5 $ with the abundance data by high resolution spectroscopy of $R > 20,000$ (right panel). 
\label{fig:barklem}\label{fig:akerman}
}
\end{figure}

\red{
In the assembly of the SAGA database, there is selection bias between the  spectroscopic studies which owe their targets to HK and HES surveys and those which rely upon the other means such as near-by high velocity stars for the target selection;  
  the former show a preference for EMP stars \citep{aoki05,aoki07,preston06,spite06,lai08,bonifacio09}, while EMP stars are scarcely found among the latter \citep{gratton00,simmerer04}. 
   The predominance of these two data is switched over around $\feoh \simeq -2.2 \mhyph -2.4$.  
}

On the other hand, some works cover a wider metallicity range. 
   \citet{barklem05} provide the carbon abundances over the metallicity range of $-3.4 \lesssim \feoh \lesssim -1.5$, though with the moderate resolution of $R = 20,000$, which account for 43 \% of all the sample stars with the carbon abundances measured.  
\red{
   Figure~\ref{fig:barklem} (left panel) shows their abundance data for the giants, which are defined by the surface gravity of $\log g \le 3.5$ and the effective temperature $T_{\rm eff} < 6000$ K (see Paper~I). 
   The \STV\ peaks at $\feoh = -2.2$ nearly by $2\sigma$ significance level despite large observation errors.  
   This fact suggests that the origin of the variation across $\feoh \simeq -2.2$ is not the systematic errors between the works biased toward EMP stars and toward stars of higher metallicity, mentioned above.
   As for the variation across $\feoh = -1.8$, the number of their data in \MP\ division is too small to give any statistically significant result.  
}
   
\citet{akerman04} also derive the carbon abundances for 28 dwarfs stars in $\feoh < -1$, distributed more or less evenly over the metallicity from \EMP\ through \MP\ divisions;  
   we define dwarfs by the surface gravity of $\log g > 3.5$ (see Paper~I). 
   Their data follow the trend of $\mabra{C}{Fe}$ increasing for lower metallicity, and the \WT\ confirms the break at $\abra{Fe}{H} \simeq -1.8$ by $2 \sigma$ significance level despite the small sample size. 
\red{
   We check the enrichment history for dwarfs, as shown in Figure~\ref{fig:akerman} (right panel).   
   Here we deal only with the abundances, measured by high resolution spectroscopy of $R > 20,000$, almost all from \citet{gratton00} and \citet{simmerer04} in addition to \citet{akerman04}. 
   The \WT\ shows the upturn across $\feoh = -1.8$ with the statistical significance of $3\sigma$.
   It is true that there are considerable differences in their abundance data, but the variations in the mean enhancement between \MP\ and \VMP\ divisions result consistent with each other within the errors.  
   This evidences that systematic offsets among different authors are canceled out in the differential variations as long as the data from each author distribute more or less evenly over the adjacent metallicity bins. 
}

In this figure, we see that a flat part extends below $\feoh \simeq -1.8$ beyond $\feoh \simeq -2.2$, different from Fig.~\ref{fig:allfe} for all the carbon-normal stars.  
   The absence of break at $\feoh \simeq -2.2$ among the dwarf samples is common to Zn and will be discussed later in \S~\ref{sec:spatial_distribution}.
   Instead, there is a gradual ascent for $\feoh \lesssim -2.4$. 
\red{
   This ascent may be related to the trend, argued by \citet{asplund05} for dwarfs, that the 1D analyses of carbon abundances from the CH molecular band gives larger abundances than those with the 3D effect. 
   The 3D correction is more important in lower metallicities, increasing by $\sim 0.2$ dex between $\feoh = -2$ and $-3$ \citep{asplund05b}. 
   In fact, the fraction of dwarfs with the carbon abundances from the CH molecular band amounts to about a half among the sample stars with $\feoh < -2.4$.  
   Such ascent is not discernible in Fig.~\ref{fig:allfe}, however, where giants dominate over the sample stars in \EMP\ division ($\sim 75$\%).  
   This may point to the difference in the 3D effect between giants and dwarfs; 
   in fact, \citet{bonifacio09} and \citet{lai08} report the enhancements larger by $\sim 0.2$ dex or more for dwarfs than for giants. 
}


\subsection{Magnesium and the $\alpha$-elements}\label{subsec:mg}

As for the $\alpha$-elements, the SAGA database collects the abundance data, measured by the high-resolution spectroscopy of resolution $R > 20,000$ for $\sim 300$ or more stars of $\feoh < -1$ without the carbon enhancement (see Table~\ref{table:elements}).  
   Such a large number of data warrant the detailed examination with the use of these high-resolution abundances only.  
   In the following analyses, we exclude outliers with peculiar abundances as well as blue metal-poor stars from \citet{preston00} (see Appendix~\ref{app:outliers} for details). 

The mean enhancement of Mg (top-middle panel in Fig.~\ref{fig:allfe}) exhibits the steepest variation between $\feoh \simeq -1.9 \mhyph -1.7$, where it decreases by $\sim 0.1$ dex toward higher metallicity. 
   The \STV{}s also mark the highest peaks exceeding $t_{2 \sigma}$ at $\feoh \simeq -1.8$ both for the 0.3 and 0.4 dex bin widths.  
\red{   The abundance data around $\feoh \simeq -1.8$ are from the sample stars, distributed more or less evenly on either side of metallicity and between giants and dwarfs \citep{fulbright00, johnson02, stephens02, gratton03, jonsell05}, so that the break is unlikely to be influenced by possible systematic errors among different authors. } 

For $\feoh \ga -1.5$, we see an upward trend of $\mabra{Mg}{Fe}$ for higher metallicity, where the \WT\ gives the next largest \STV{}s. 
   This is influenced by the data from \citet{gratton03}, which are systematically higher than those from others, probably due to the different oscillator strengths adopted, as stated in the paper.   
   Removing their data largely alleviates the upward trend, as shown by thick dotted line, but note that the upward trend exists among their own data.
   There is another feature around $\abra{Fe}{H} \le -2.4$, where $\mabra{Mg}{Fe}$ decreases for lower metallicity and the \STV\ is elevated. 
  If limited to giants, the mean enhancement remains nearly flat below the break (dotted line for $\feoh < -3$);  
   a dip still persists around $\feoh \simeq -2.6$, which is caused by several giants with considerably small enhancements of $\abra{Mg}{Fe} \la 0.2$, swarmed near this metallicity. 

The other $\alpha$-elements, Si, Ca and Ti, bear the same characteristics as Mg (see Appendix~\ref{app:analyzed}); 
   they all exhibit the break at $\feoh \simeq -1.8$ with decrease by $\sim 0.1$ dex for higher metallicity, and lack the break at $\feoh \simeq -2.2$.  
   They also share some feature above $\feoh > -1.5$ and undulations for $\feoh \la -2.5$ with low statistical significance.   


\subsection{Zinc and the iron-peak elements}\label{subsec:Zn} 

Among the three iron-peak elements in Fig.~\ref{fig:allfe}, Ni and two others, Co and Zn, exhibit different behaviors. 
   The former shares the features with the $\alpha$-elements (see Appendix~\ref{appsubsub:Ni}). 
   On the other hand, the latter two display distinctive behaviors, i.e., the greatest ascent of mean enhancement for lower metallicity and an additional break at the lowest metallicity. 
   Thus, Zn is endowed with all the three breaks that we find, while Co lacks the highest metallicity break through the scarcity of data in \MP\ division (see Appendix~\ref{appsubsub:Co}).  


\begin{figure}
\begin{center}
\includegraphics[width=95mm,scale=1,bb=92 40 590 305, clip]{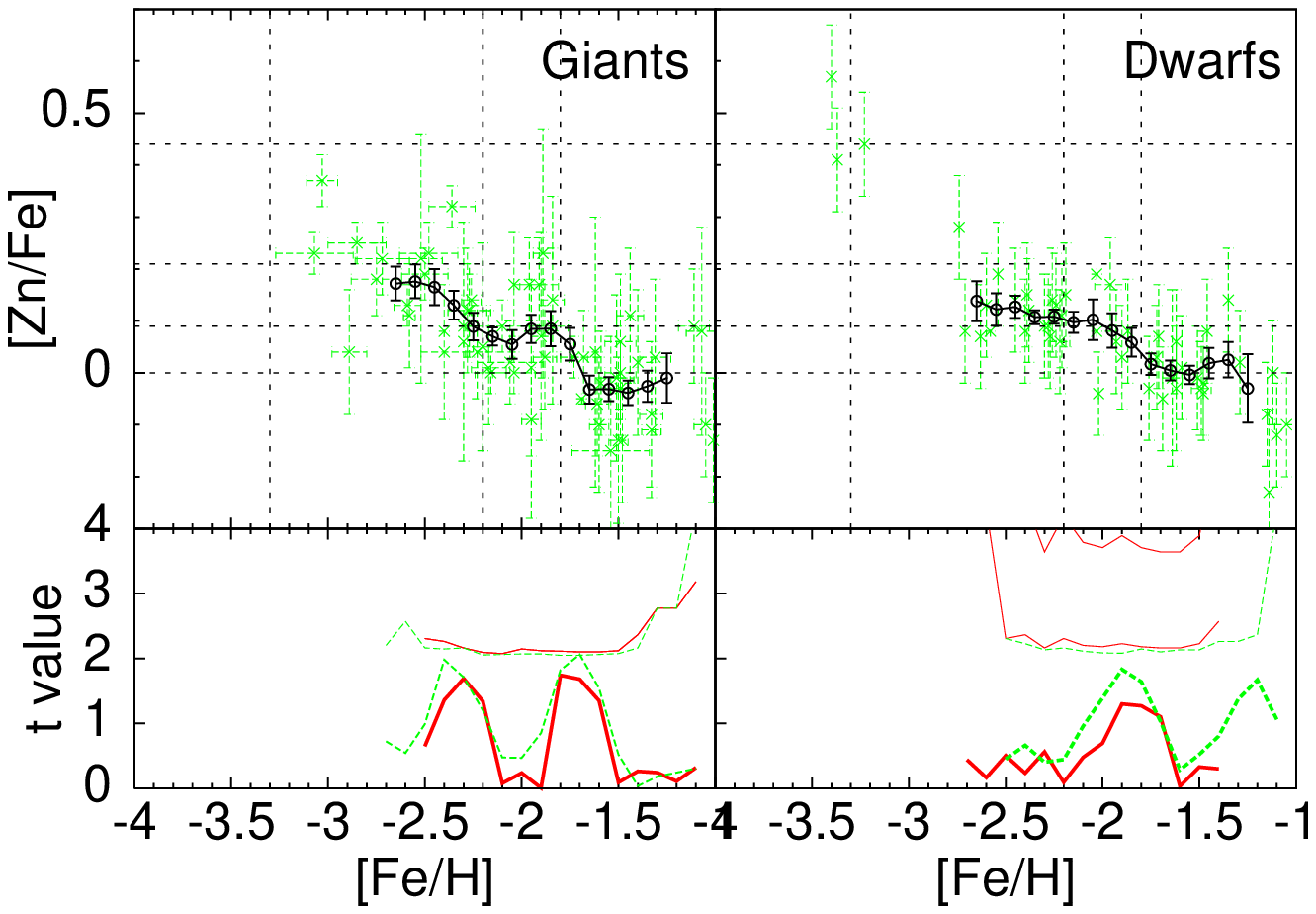}
\end{center}
\caption{The same as Fig.~\ref{fig:allfe} but for giants (left panel) and dwarfs (right panel) with the Zn abundances, from \citet{mishenina02}, \citet{nissen02}, \citet{nissen04}, \citet{johnson02} and \citet{saito09}. } \label{fig:znfe2}
\end{figure}


In the enrichment history for Zn in Fig.~\ref{fig:allfe}, we may recognize such features as pointed out by the previous works, i.e., overall increase for lower metallicity and discontinuous change around $\feoh \simeq -2$, as mentioned in \S 1.   
   The mean enhancement increases from $\mabra{Zn}{Fe} \simeq 0.0$ at $\feoh \simeq -1.7$ to 0.5 at the lowest metallicity, which is almost the same as Co. 
   And yet, the increase is not constant and nearly flat parts are discernible midway, as is also the case for Co.  
   Indeed, the \WT\ reveals three peaks of \STV\, reaching to $t_{2 \sigma}$ for the 0.3 dex bin width.   
   The depression of \STV\ between the two metal-rich peaks at $\feoh = -1.8$ and $-2.2$ grows deeper for smaller bin width, indicative of the in-between flat distribution of mean enhancement, as in the case for carbon. 
   The lowest metallicity break takes place at $\feoh = -3.2$, almost the same metallicity, but slightly smaller than $\feoh = -3.3$ for Co, the difference of which will be discussed below in \S~\ref{subsec:vertical_variation}.  
   In this connection, it should be noted that an upturn of $\mabra{Zn}{Fe}$ for $\feoh < - 2.8$ is an artifact of observational bias.  
   In the sample stars, the Zn abundances are mostly bounded by $\abra{Zn}{H} \ga -2.9$ (dotted line in Fig.~\ref{fig:allfe}) except for \citet{cayrel04}, who observe down to $\abra{Zn}{H} \simeq -3.5$. 
   The scarcity of data below this abundance gives rise to a superficial increase of the mean enhancement for lower metallicity when averaged over a given metallicity bin.  

\red{
For Zn, several works cover a fairly large number of sample stars, distributed more or less evenly in a wide range of $\feoh \simeq -3 \mhyph -1$ \citep[][see Table~\ref{table:elements}]{johnson02,mishenina02,nissen07,saito09}.  
   We examine the enrichment history for the data from these works only in order to minimize possible effects of systematic errors among different authors.  
   Figure~\ref{fig:znfe2} shows the results for giants and dwarfs separately. 
   Giants show basically the same step-wise distribution of mean enhancements as in Fig.~\ref{fig:allfe};  
   the \STV{}s have two peaks at nearly the same metallicities of $\feoh = -1.8$ and $-2.3$, though the statistical significance is lower because of smaller sample size.  
   In contrast, the mean enhancement of dwarfs present only the feature at $\feoh \simeq -1.8$ and remain nearly flat down to $\feoh \simeq -2.7$.   
   We return to the absence of break at $\feoh \simeq -2.2$ for dwarfs and also to the slightly lower break metallicity and mean enhancement in \EMP\ division for giants when discussing the spatial distributions in \S~\ref{sec:spatial_distribution} 
}

The results indicate that the variation of mean enrichment is not steady nor continuous but a term of constant or slowly changing $\mabra{Zn}{Fe}$ is replaced by another one at the breaks.  
   Thence, we may assume step-wise variations with sharp differences in narrow metallicity ranges across the three break metallicities, as discussed for C above. 
   The similar step-function like change is seen in Fig.~4 of \citet{saito09} as a leap at $\feoh = -2$ in the diagram of mean enhancement at intervals of 0.5 dex;   
   in their figure, $\mabra{Zn}{Fe}$ rises continuously below $\feoh = -3$ also for the same reason as discussed above.

\section{Overall picture of enrichment history}\label{sec:picture}


\begin{table*}
\begin{center}
\caption{The characteristics of element enhancements for the four stellar populations among the sample stars in the SAGA database.}
\vskip 5pt 
\label{table:average}
\label{table:sigma}
\begin{tabular}{ccccc}
\hline
Divisions & \MP({\MP}2) & \VMP\ & \EMP\ & \UMP\ \\
& $-1(-1.5) \ge \feoh > -1.8$ & $ -1.8 \ge \feoh > -2.2$ & $-2.2 \ge \feoh > -3.3$ & $-3.3 \ge \feoh $ \\
\hline
Elements & \multicolumn{4}{c}{average enhancement} \\
\hline
C & $-0.11 \pm 0.03$ & $0.03 \pm 0.06$ & $0.31 \pm 0.03$ & $0.27 \pm 0.11$ \\
Mg & $0.36 \pm 0.02$ ($0.33 \pm 0.03$) & $0.42 \pm 0.02$ & $0.38 \pm 0.02$ & $0.34 \pm 0.04$ \\ 
Si & $0.31 \pm 0.02$ & $0.39 \pm 0.03$ & $0.45 \pm 0.03$ & $0.43 \pm 0.05$ \\ 
Ca & $0.29 \pm 0.01$ & $0.36 \pm 0.02$  & $0.34 \pm 0.01$ & $0.30 \pm 0.04$ \\ 
Ti II & $0.26 \pm 0.01$ & $0.31 \pm 0.03$  & $0.33 \pm 0.01$ & $0.32 \pm 0.04$ \\
Sc & - & $0.08 \pm 0.04$ & $0.16 \pm 0.02$ & $0.16 \pm 0.04$ \\
Co & - &$0.12 \pm 0.06$ & $0.28 \pm 0.02$ & $0.45 \pm 0.04$ \\
Ni & $-0.05 \pm 0.01$ ($-0.07 \pm 0.02$) & $0.01 \pm 0.03$ & $0.00 \pm 0.02$ & $0.00 \pm 0.03$ \\
Zn & $0.00 \pm 0.02$ & $0.09 \pm 0.03$ & $0.22 \pm 0.02$ & $0.44 \pm 0.04$ \\
\hline
Elements & \multicolumn{4}{c}{scatter of distribution (standard deviation $\sigma$) } \\
\hline
C & $0.16 \pm 0.03$ & $0.20 \pm 0.03$ & $0.19 \pm 0.02$ & $0.24 \pm 0.05$ \\
Mg  	& $0.14 \pm 0.02$ & $0.13 \pm 0.01$ & $0.17 \pm 0.01$ & $0.19 \pm 0.03$ \\
Si  	& $0.14 \pm 0.01$ & $0.14 \pm 0.02$ & $0.15 \pm 0.02$ & $0.14 \pm 0.04$ \\ 
Ca  	& $0.09 \pm 0.01$ & $0.07 \pm 0.01$ & $0.11 \pm 0.01$ & $0.19 \pm 0.06$ \\ 
Ti II	& $0.08 \pm 0.01$ & $0.10 \pm 0.02$ & $0.12 \pm 0.02$ & $0.12 \pm 0.02$ \\
Sc	&   -            &  $0.06 \pm 0.01$  &  $0.16 \pm 0.01$  &  $0.23 \pm0.03$  \\ 
Co	&  -             &  $0.05 \pm 0.02$  &  $0.14 \pm 0.01$  &  $0.18 \pm 0.04$  \\ 
Ni 	& $0.08 \pm 0.01$  &  $0.06 \pm 0.01$  &  $0.09 \pm 0.01$   &  $0.17 \pm 0.02$  \\
Zn	& $0.08 \pm 0.01$  &  $0.07 \pm 0.01$  &  $0.14 \pm 0.03$  &  $0.06 \pm 0.03$  \\
\hline
\end{tabular}
\end{center}
\end{table*}


\begin{figure}
\begin{center}
\includegraphics[width=95mm,scale=1,bb=70 50 390 418, clip]{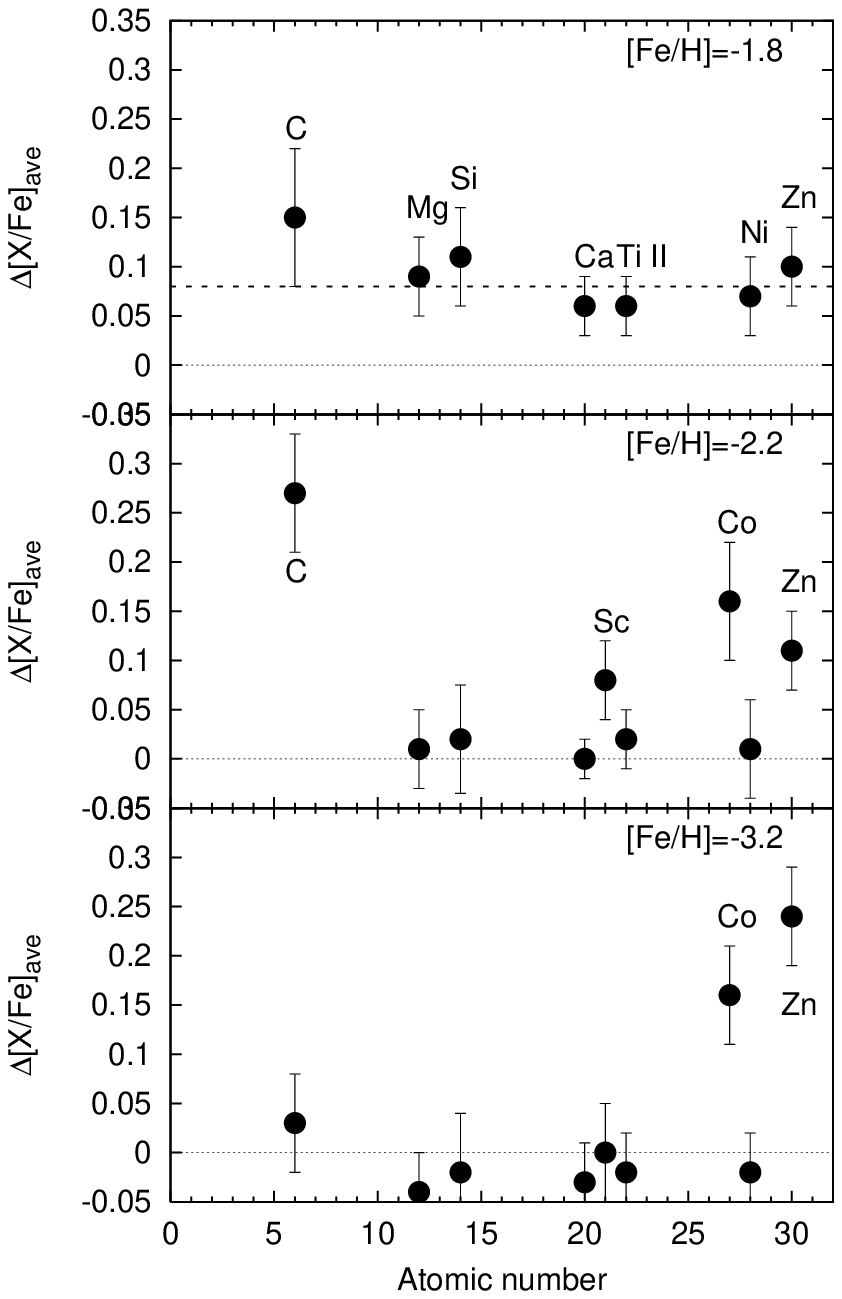} 
\end{center}
\caption {Differences in the average enhancements, $\Delta \avabra{X}{Fe}$, of the sample stars between the successive metallicity divisions, partitioned by the breaks.
   In top panel, \MP\ division is limited to $\feoh < -1.5$ to avoid an upturn trend in higher metallicity, while \VMP\ division is extended down to $\feoh = -2.5$ for $\alpha$-elements and Ni, devoid of the break at $\feoh = -2.2$. 
   Dashed line denotes the weighted average of differences. 
   In middle panel, the metallicity range of \EMP\ division is limited to $\feoh > -2.5$ to avoid the dwarf/giant discrepancy (see the text). 
\label{fig:dxfe}}
\end{figure}

\subsection{Four stellar populations of different abundance patterns}\label{subsec:four_populations}

We have revealed the sharp variations, or breaks, of mean enhancements across three metallicities with the statistical significance of $2 \sigma$ level or more among the sample stars.  
\red{
   The breaks occur for multiple elements and the metallicities coincide within the errors.  
   Furthermore, the presence or absence of breaks is common to the elements that share the nucleosynthetic characteristics but not the atomic properties.  
   These facts make it unlikely that the breaks happen to be caused by the selection bias and/or systematic errors, possibly involved in the assembly of data from the different authors.  
}

From the \WT, we may also regard the mean enhancement as constant between and outside the breaks.  
   Table~\ref{table:average} lists the average enhancements, $\avabra{X}{Fe}$, of sample stars in the four metallicity divisions, partitioned by the breaks, for all the elements studied.  
   The break metallicities are set at $\feoh = -1.8$, $-2.2$ and $-3.3$, the last one from Co (see below in \S~\ref{subsec:vertical_variation}). 
   These averages define the abundance patterns for the stellar populations that dominate over the sample stars in each metallicity division. 
   We call the stars with these abundance patterns, Pop~IIa, Pop~IIb, Pop~IIc and Pop~IId, in order of increasing metallicity.

The variations of average enhancements are all downturns for higher metallicity but different in their characteristics as shown in Figure~\ref{fig:dxfe}.   
   The break at $\feoh = -1.8$ is detected for all the elements except for Co and Sc which lack the data in \MP\ division. 
   In addition, the rates of variation agree within the errors to give the weighted average, $\Delta \avabra{X}{Fe} = 0.08 \pm 0.04$. 
   Accordingly the decline in the mean enhancement is readily explained in terms of an increase in the Fe yield rather than the decreases of all other elements by almost the same ratio to iron.  
   In contrast, the variations of $\avabra{X}{Fe}$ across the other two breaks differ from elements to elements.  
   For the break at $\feoh = -2.2$, the variations are observed for C, two iron-peak elements, Co and Zn, and possibly Sc, but not for all the $\alpha$-elements and an iron-peak element Ni.  
   For the most metal-poor break at $\feoh = -3.3$, only Co and Zn display the statistically significant differences.  

Accordingly, the four populations have one-to-one correspondence to the average enhancements of Zn and probably of Co;  the very high ($\avabra{Zn}{Fe} = 0.44$), high (0.22), low (0.09), and very low (0.00) for Pops~IIa, IIb, IIc, and IId, respectively (see \S\ref{subsec:vertical_variation}. 
   In contrast, there is only one break for $\alpha$-elements and Ni.  
   Such a diversity of variations among the elements suggests different mechanisms of breaks, as discussed later in \S~\ref{sec:origin}.

\subsection{Variations in the scatters of enhancement  distributions }\label{subsec:enhancement_distributions}

In Table~\ref{table:sigma}, we also show the standard deviation, $\sigma$, of the Gaussian fit to the distribution of enhancements in each metallicity division, as shown in Fig.~\ref{fig:hist-cfe} for carbon. 
   The scatters of the distribution are directly affected by systematic errors among the different authors if exist, and yet, they have some interesting statistical trends which may also carry information on the  supernova yields. 

Between \VMP\ and \MP\ divisions, the scatters remain unchanged within errors for all the elements despite the peak shifts.  
   In particular, the heavier elements from Ca through the iron-peak elements display a rather narrow distribution with the scatters as small as typical observation errors. 
   It makes a contrast with the three lighter elements, C, Mg, and Si, which have much larger scatters. 
   These elements are distinguished in that the latter are synthesized during the quasi-static evolution and form the shells outside the iron core prior to SN explosions, whereas the former are synthesized only in the iron core.  
   Accordingly, the larger scatters of three lighter elements may be related to the variations in the amount of these elements that preexist outside the iron core and can be saved from the processing by explosive burning.   

For \EMP/\UMP\ divisions, there is a general tendency of the scatters increasing for lower metallicity, possibly due to larger observational and/or systematic errors among the authors.  
   We note, however, that the increase from \VMP\ to \EMP\ division is larger for the elements with the peak shift, except for C which has sufficiently large scatter already in \VMP\ division.  
   This suggests that the peak shift correlats with the variation of scatter between \EMP\ and \VMP\ divisions, differently from between \VMP\ and \MP\ divisions. 


\section{Correlation with spatial distributions} \label{sec:spatial_distribution}

\subsection{Spatial distribution of sample stars} 

\begin{figure}
\begin{center}
\includegraphics[width=110mm]{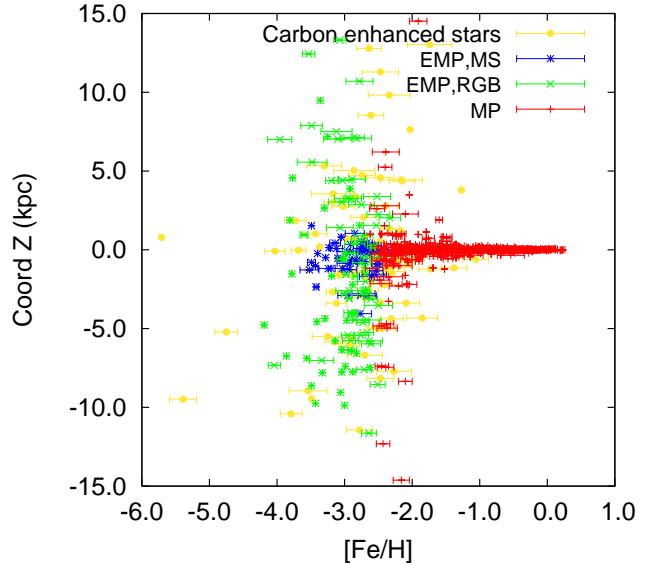}
\end{center}
\caption{
Distribution of the Galactic coordinate, $Z$, perpendicular to the Galactic plane for all the sample stars with the abundances by high resolution spectroscopy of resolution $R > 20,000$ in the SAGA database, plotted against the metallicity.  
   In the legend, MP and EMP denote the stars of $\feoh > \hbox{ and } \le -2.5$ and RGB and MS denote giants with the surface gravity $\log g \le 3.5$ and the effective temperature $T_{\rm eff} \le 6000$ K and other than giants, i.e., dwarfs and a very few blue HB stars, respectively, for the sample stars without the carbon enhancement.
\label{fig:Zfe} }
\end{figure}

\begin{figure}
\begin{center}
\includegraphics[width=80mm]{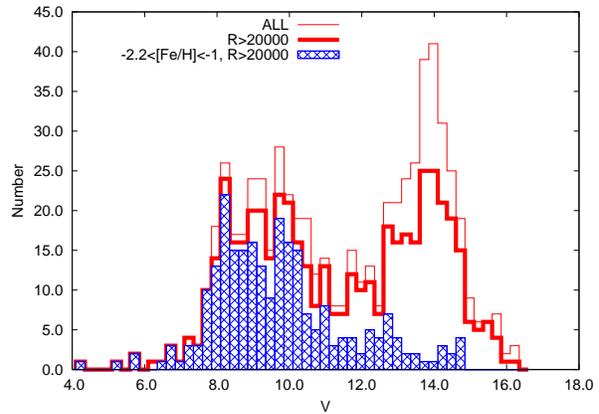} 
\end{center}
\caption{Distribution of $V$ magnitude of sample stars without the carbon enhancement in the SAGA database (thin line).  
   Thick lines denote the sample stars with the abundance data by high-resolution spectroscopy of $R > 20,000$, and hatch represents the stars of $\feoh >-2.2$.  
\label{fig:vmag}}  
\end{figure}

\begin{figure}
\begin{center}
\includegraphics[width=90mm,scale=1,bb=20 42 380 335, clip]{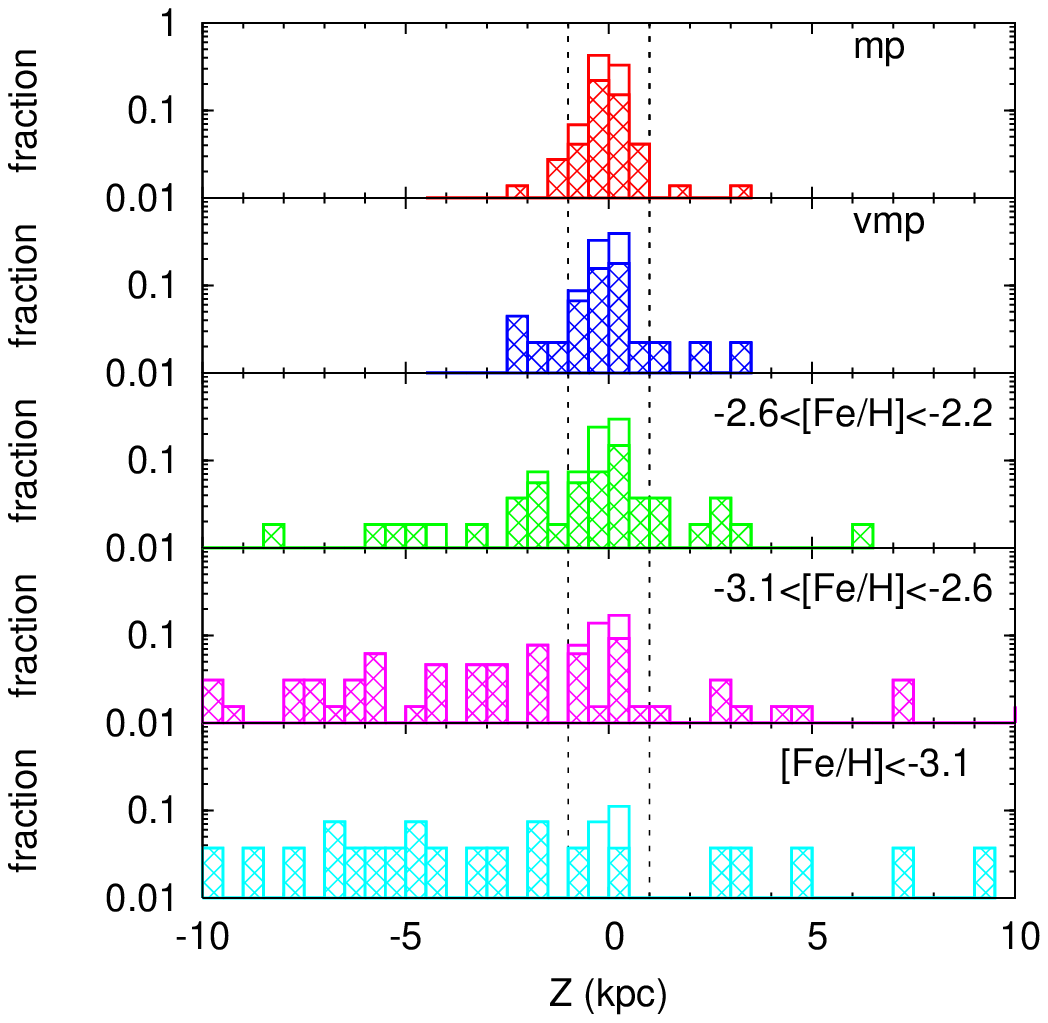}
\end{center}
\caption{
   Variations of the distribution along the Galactic $Z$ coordinate with the metallicity for the sample stars with the Zn abundances by high-resolution spectroscopy. 
   Histogram denotes the normalized number of sample stars with 500 pc bin and hatch indicates the fraction of giants.   
}
\label{fig:two_compZ}
\end{figure}

The SAGA database provides the spatial distributions of sample stars, evaluated with the use of the spectroscopic distances (see Paper~I).  
   Figure~\ref{fig:Zfe} shows the distribution of sample stars along the Galactic coordinate, $Z$, perpendicular to the Galactic plane against the metallicity.
There is a marked difference in the spatial distributions according to the metallicity.   
   The sample stars spread over the whole Galactic halo through $|Z| \simeq 15$ kpc below $\feoh \simeq -2.2$, while above it, they concentrate near the Galactic plane. 
   The carbon-enhanced stars are observed all over the Galactic halo for slightly higher metallicity of $\feoh \simeq -2.0$.  
   These distinct distributions result from the target selections for the spectroscopic observations.  
   We show the $V$ magnitude distribution of sample stars in Figure~\ref{fig:vmag}, which is bimodal with the peaks at $V \simeq 8 \mhyph 10$ and $V \simeq 13 \mhyph 15$ (see also Fig.~6 of Paper~I).   
   The fainter peak is composed of the stars from the works, which take the targets from HK and HES surveys in favor of stars in \EMP\ and \UMP\ divisions.  
   On the other hand, the brighter peak is composed mostly of stars in \VMP\ and \MP\ divisions, taken from the works which observe nearby targets such as selected based on the kinematics.  

Figure~\ref{fig:two_compZ} illustrates the variations of spatial distributions with the different metallicity divisons for the sample stars with the Zn abundances by the high-resolution spectroscopy.   
   The distribution is narrow with the scaleheight of $|Z| \la 0.5$ kpc in \MP\ and \VMP\ divisions, and extends nearly flat through $\sim 10$ kpc in \EMP/\UMP\ divisions.  
   In addition, for $-2.6 \le \feoh <-2.2$, we see that stars with $|Z| \la 0.5$ kpc stand out above the extended distribution;  
   this concentrated component decreases relative to the extended component for lower metallicity and grows indiscernible for $\feoh < -3.1$. 
\red{   The ratio of two components is subject to the selection bias of observed targets, as stated above, and yet, the presence of concentrated component in a part of \EMP\ division and its absence for still lower metallicity may reflect the actual distributions since the precise metallicity of these stars is known only after the high-dispersion spectroscopic observations. 
}

\subsection{Variation of the abundance patters along the height from the Galactic plane} \label{subsec:vertical_variation}


\begin{figure*}
\begin{center}
\includegraphics[width=170mm,scale=1,bb=0 35 580 475, clip]{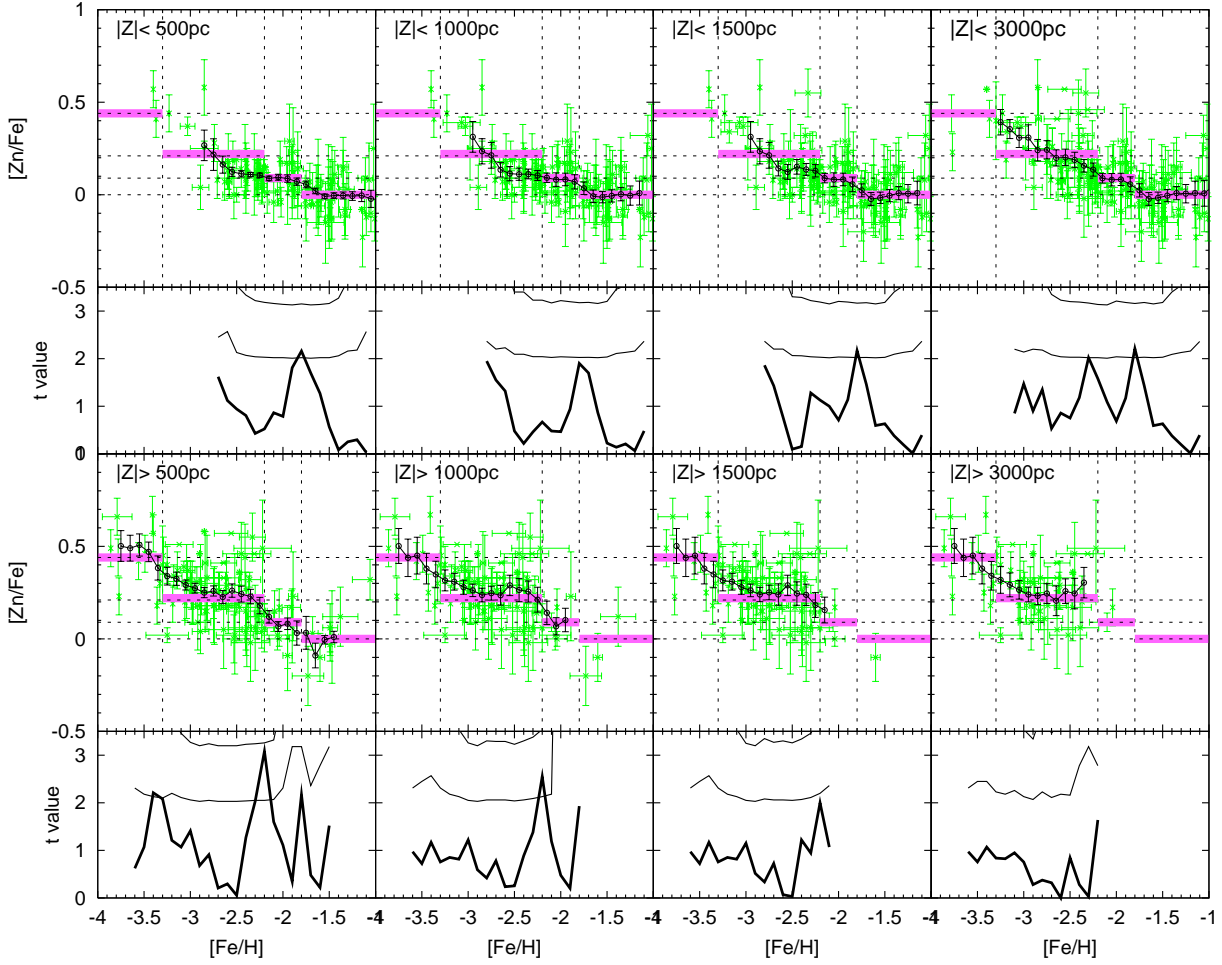}
\end{center}
\caption{
Variations of Zn enhancement with the height, $|Z|$, from the Galactic plane. 
  Upper and lower panels show the enrichment history of Zn and the results of \WT\ for the nearby and distant subsets of sample stars, respectively, with the dividing height increasing from left to right as given at the top of each panel. 
   Vertical and colored horizontal lines denote the break metallicities and the average enhancements of Pops~IIa, IIb, IIc, and IId), respectively.  
\label{fig:znfesp} }
\end{figure*}

Figure~\ref{fig:znfesp} compares the Zn enhancements for the nearby and distant subsets of sample stars, divided by the height, $|Z|$, from the Galactic plane.  
   We have defined the four stellar populations as dominating over the sample stars in each metallicity division with the one-to-one correspondence to the average enhancements of Zn.  
   This figure demonstrates, however, that the distribution of mean enhancements varies with the height from the Galactic plane.  

For the nearby subset with the height $|Z| \le 0.5$ kpc, the mean enhancement exhibits the ascent from the  Pop~IId to the Pop~IIc averages across $\feoh \simeq -1.8$, and then, remains there until it increases again for $\feoh < -2.6$ without undergoing the ascent to the Pop~IIb average at $\feoh \simeq -2.2$.  
   For the distant subset of $|Z| > 0.5$ kpc, on the other hand, $\mabra{Zn}{Fe}$ traces the same variation as in Fig.~\ref{fig:allfe}, including the upturn from the Pop~IIc to Pop~IIb averages across $\feoh \simeq -2.2$.  
   The \WT\ shows that the break at $\feoh \simeq -2.2$ stands out with nearly $3 \sigma$ level for the distant subset but no sign for the nearby subset. 
   In addition, for the distant subset, the lowest-metallicity break shifts to slightly lower metallicity than in Fig.~\ref{fig:allfe} to be same as that of Co, as a result of removal of the nearby sample stars.  
   This leads us a conjecture upon different spatial distributions for Pop~IIa and Pop~IIb, though the sample size is too small for the statistics.  

As the dividing height increases, the trends change both in the nearby and distant subsets. 
   For the nearby subsets, the mean enhancement in \EMP\ division augments and approaches the average of Pop~IIb. 
   Eventually, the break appears at $\feoh \simeq -2.3$ for the dividing height $|Z| \simeq 3$ kpc with the $2 \sigma$ statistical significance level. 
   The distribution of $\mabra{Zn}{Fe}$ are left unchanged in \VMP\ and \MP\ divisions.  
   For the distant subsets, on the other hand, the distribution remains little changed in \EMP\ and \UMP\ divisions, but Pop~IId and Pop~IIc are made indiscernible for $|Z| > 1$ and 1.5 kpc because of the lack of sample stars for $\feoh > -2.2$.   

Accordingly, Pop~IIc resides with Pop~IIb over a wide range of metallicity in \EMP\ division, segregated spatially. 
   The coexistence starts to be discernible from $\feoh \simeq -2.6$, and the fraction of Pop~IIc stars increases for higher metallicity.  
\red{   This may explain the absence of the break at $\feoh \simeq -2.2$ for dwarfs in Fig.~\ref{fig:znfe2} since all of the dwarfs in \EMP\ division fall below $|Z| = 500$ pc.   
   The slight smaller break metallicity and mean enhancement in \EMP\ division for giants are also attributable to the fraction of the nearby giants which increases toward higher metallicity.  
}

The above result urges us to reconsider the higher metallicity end of Pop~IIb as well.  
   Among the sample stars, the number of distant stars sharply decrease for $\feoh > -2.2$ and few stars have the Zn abundance measured for $|Z| \ga 3$ kpc in \VMP/\MP\ divisions, as seen in Figs.~\ref{fig:Zfe} and \ref{fig:two_compZ}.  
   It is likely, however, that Pop~IIb stars persist beyond $\feoh = -2.2$ in the extended halo, as inferred from the distribution of carbon-enhanced stars.  
   It waits for future observations of distant stars to determine the extension and properties of Pop~IIb in higher metallicity.   

In this section, we have dealt with the Zn enrichment only.  
   There are three more elements, C, Co, and possibly Sc, with the different average enhancements between Pops~IIb and ~IIc.  
   They lack sufficient abundance data for nearby stars in \EMP\ division, however.  
   We should wait for future observations, in particular, such works covering a wide range of metallicity as done for Zn.   

\subsection{Connection to the kinematically resolved components}\label{subsec:vphi}  

The Galactic halo is known to be kinematically complex and comprise of multiple components \citep[e.g.,][]{carney84,chiba00,lee07,miceli08}.
   Recently, \citet{carollo10} derive the kinematic properties of the outer and inner halos (OH and IH), the metal-weak thick disk (MWTH), and the thick disk (TD) from the data of the Slone Digital Sky Survey.  
   For the nearby sample stars, the kinematical data are given in the papers \citep{fulbright02,stephens02,gratton03,ishigaki10,schuster06,schuster12,silva12}.   
   We may use these data to discuss the relationship of the stellar populations, derived above, to the kinematically resolved components.  


\begin{figure}
\begin{center}
\includegraphics[width=95mm,height=60mm,bb=50 50 500 302, clip]{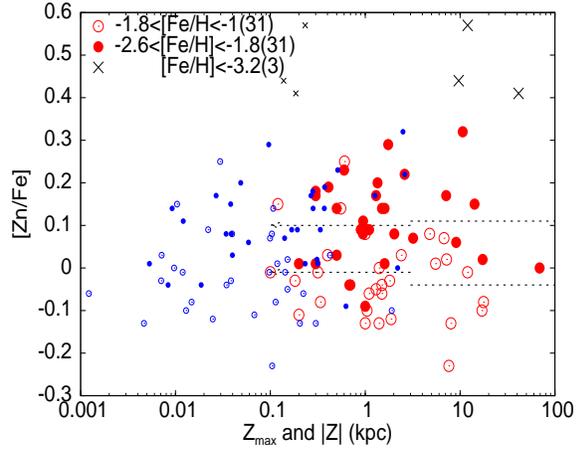}
\end{center}
\caption{Enhancement of Zn, plotted against the maximum height, $\zmax$, reached by their orbits (large symbols), and the observed height, $|Z|$ (small symbols), from the Galactic plane for the sample stars with the kinematic data available in the literature. 
   The mean Zn enhancements for the sample stars in \VMP\ and \MP\ divisions are plotted by upper and lower broken lines, respectively, derived separately for $\zmax \le $ and $> 3$ kpc.  
}
\label{fig:znzmax}
\end{figure}
   
Figure~\ref{fig:znzmax} shows the enhancement of Zn as a function of the maximum height, $\zmax$, of their orbits for 65 sample stars with the kinematic data available. 
  For the stars in \VMP/\EMP\ and \MP\ divisions, the distributions of $\zmax$ result similar with 77 and 71\% stars below $\zmax < 3$ kpc, respectively.  
   Their mean enhancements agree with the averages of Pop~IIc and Pop~IId within the errors, respectively.  
   In adition, $\zmax$ is available for 3 stars of $\feoh < -3.2$, which have $\zmax \ga 10$ kpc and the enhancements as large as the Pop~IIa average.  
   For other elements (Mg, Si, Ca, Ti II and Ni), the distributions of $\zmax$ are almost similar  
   with $\sim 70$\% stars at $\zmax < 3$ kpc 
   both for \VMP\ and \MP\ divisions. 



\begin{figure}
\begin{center}
\includegraphics[width=95mm,height=55mm,bb= 0 20 650 302]{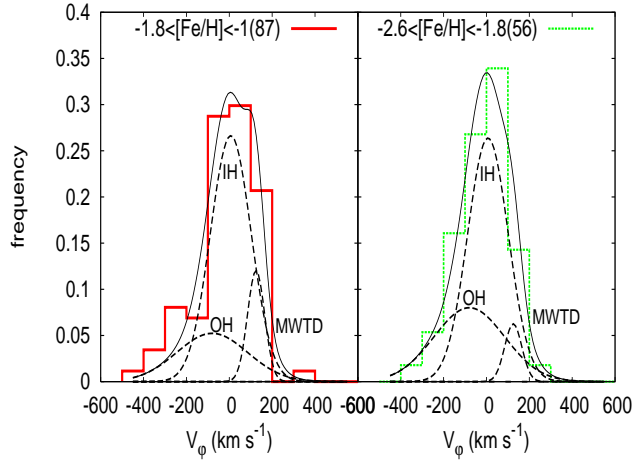}
\end{center}
\caption{
The distributions of rotation velocity, $V_\phi$, around the Galactic rotation axis for sample stars of $|Z| < 1$ kpc with the Zn and/or Mg abundances in \MP\ and \VMP/\EMP\ divisions, respectively. 
   Thick and broken lines illustrate the Gaussian decomposition with the use of the kinematically resolved components, the outer halo (OH), the inner halo (IH), and the metal-weak thick disk (MWTD) by \citet{carollo10}.   
}
\label{fig:rotdist}
\end{figure}

Figure~\ref{fig:rotdist} illustrates the distributions of rotation velocity, $V_\phi$, for the nearby sample stars of Pop~IIc and Pop~IId.  
   Both populations display almost the same distributions with the peak, $V_\phi = 2.7 \pm 9.1$ and $32 \pm 18 \hbox{ km sec}^{-1}$, and the standard deviation, $\sigma = 120$ and $115 \hbox{ km sec}^{-1}$, respectively.   
   Slight shift of the peak is discernible but it reduces if we limit the metallicity range to $\feoh < -1.5$.  
\red{ 
   In this figure, we superpose the decompositions of $V_\phi$ distributions into those of kinematically resolved components, taken from \citet{carollo10}. 
   The contribution of the inner halo is overwhelming for both populations (62 and 65\% for Pop~IIc and Pop~IId, respectively). 
   The metal-weak thick disk increases the contribution with the metallicity (6 and 13\%), while the thick disk makes negligible contribution.  
   On the other hand, the contribution from the outer halo amounts relatively large for the both populations (32 and 22\%), which forms a contrast to the argument by \citet{carollo10} that the IH component dominates over the OH components up to $\zmax \simeq 10 \mhyph 15$ kpc. 
   This may be attributable to the selection bias in our sample stars by the proper motion and/or radial velocity, which take up stars with kinematically larger deviations.    
}

\red{
In summary, Pops~IIc and IId are kinematically consecutive and largely overlap with the inner halo.  
  On the other hand, judging from their extended spatial distributions, we may well consider Pops~IIa and IIb belonging to the outer halo, whose metallicity distribution function peaks at $\feoh \approx -2.2$ \citep{carollo07}. 
}

\section{Discussion: The star formation history}%
 \label{sec:origin}


\begin{figure}
\begin{center}
\includegraphics[width=95mm,height=50mm,bb= 0 0 820 450]{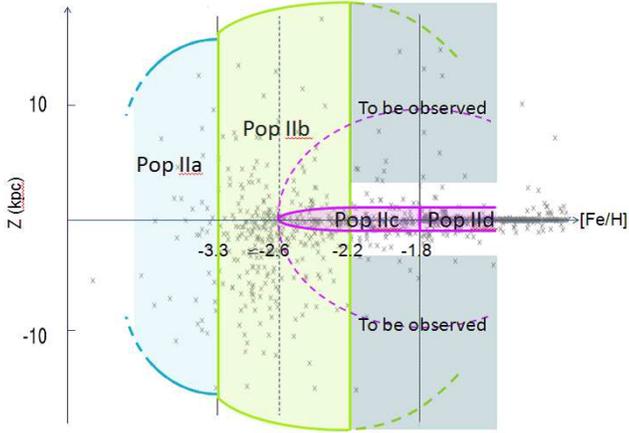}
\end{center}
\caption{
Schematic drawing of the domains of the four stellar populations on the metallicity and spatial distribution diagram.  
   In the background plotted are the sample stars, assembled in the SAGA database, and broken lines suggest the possible extensions of their domains.   
}
\label{fig:domain}
\end{figure}

We have identified the four stellar populations among the low-mass stars in the Galactic halo, different not only in the metallicity and abundance pattern but also in the spatial distribution.    
   Figure~\ref{fig:domain} illustrates their domains on the diagram of the metallicity and the height from the Galactic plane.  
   Two lower-metallicity populations, IIa and IIb, spread over the halo and form the substructures of the outer halo.  
   On the other hand, two higher-metallicity populations, IIc and IId, are kinematically consecutive and overlap the inner halo. 
   In addition, two middle-metallicity populations, IIb and IIc, coexist in a wide range of metallicity, segregated by the spatial distributions.  
   In this section, we discuss the origins and relationship of these stellar populations from the differences in the averaged abundance patterns and the spatial distributions.  

   The different patterns of average enhancements may stem from the dependences of supernova yields on the metallicity and mass of progenitors.  
   The mass dependence of SN yields manifests itself through the initial mass function (IMF) of progenitors.  
   The slope of IMF over the mass range of CCSNe affect the enhancement distribution, which may vary with the metallicity and also with the physical conditions of star forming sites.    
   In addition, the averaged yields change with the elapsed time of star formation because of large difference in the evolutionary time to supernovae between CCSNe and Type Ia SNe, which introduces the timescale into the enrichment history and gives information on the star formation rate (SFR). 
   The abundance patterns of stellar populations and their variations may, therefore, impose constraints not only on the supernova yields but also on the IMF and SFR of progenitors, through which we may gain an insight into the formation process of the Galactic halo.  
   In the following, we start with the examination of the abundance patterns and advance to discuss the possible sites of these stellar populations during the Galaxy formation.  

\subsection{Transition between two higher-metallicity populations and the star formation rate }\label{subsec:SFR} 

A salient feature of the $\feoh \simeq -1.8$ break is that the differences of $\avabra{X}{Fe}$ agree with each other within errors for all the elements studied.  
   It is readily to be interpreted, therefore, in terms of the increase of iron production rather than decreases in the production of all other elements.  
   Such a surplus of iron yield over the other elements is provided by Type Ia SNe that explode with the evolutionary time much longer time than CCSNe. 
   
If this is the case, Pop~IIc stars has to raise the metallicity up to $\feoh \simeq -1.8$ by CCSNe before the delayed iron production by Type Ia SNe comes to be dominant. 
   This condition places a constraint on the average star formation rate (SFR) of Pop~IIc stars;  
\begin{eqnarray} 
\left. \frac{ d \log M}{d t} \right |_{\rm Pop~IIc} & \simeq & 4 \times 10^{-11} \hbox{ yr}^{-1} \times \left( \frac{\tau_{\rm delay}}{10^9 \hbox{ yr}} \right)^{-1}\nonumber \\
 & & \left( \frac{M_{\rm Fe}}{0.07 \msun} \right) ^{-1} \left( \frac{n_{\rm CCSNe}}{0.007 \msun^{-1}} \right)^{-1}, 
\label{eq:sfr-VMP}
\end{eqnarray}  
   where $\tau_{\rm delay}$ is the delay time for Type Ia SNe to dominate over the iron production:  
   $M_{\rm Fe}$ and $n_{\rm CCSNe}$ are the average iron yield of CCSNe and their number per unit mass incorporated into the stars at the formation.   
   Here we take the delay time at $\tau_{\rm delay} = 1$ Gyr and the mass range of CCSNe at $M = 10 \mhyph 50 \msun$.  
   As for the IMF, we assume the Salpeter IMF with the lower mass limit $M_l \simeq 0.33 \msun$ (the dependence on $M_l$ is given by $n_{\rm CCSNe} \propto M_l{}^{-0.35}$).   
   For the Pop~IIc population, therefore, the specific SFR is rather small, and comparable to the SFR ($\sim 10^{-10} \hbox{ yr}^{-1}$) in the present Galaxy \citep[e.g., the review by][]{kennicutt12}.  

A peculiarity of this transition is that $\mabra{X}{Fe}$ displays decrease only by $\sim 0.1$ dex, however.  
   It is in contrast to the contributions of Type Ia SNe in more metal-rich regime of $\feoh > -1$ \citep[e.g. see][]{tsujimoto95} and in some of dwarf galaxies \citep[][]{tolstoy09}, which reduce $\mabra{\alpha}{Fe}$ by several tenth dex down to the solar value or even below.  
   This indicates either that the star formation of Pop~IId is interrupted on the way of decreasing $\mabra{X}{Fe}$, or contrariwise that the specific SFR is sufficiently elevated. 
   In the latter case, the yields of CCSNe overwhelm those from Type Ia SNe again to cease the decrease of $\mabra{X}{Fe}$ or even switch to an upturn.  
   In actuality, the thick disk is ascribed to the star burst, which are aged $12 \mhyph 13$ Gyr and raise the average metallicity up to $\feoh \simeq -0.6$ \citep[e.g.,][]{fuhrmann11}; 
   the specific SFR is estimated to be larger by an order of magnitude  
\begin{equation} 
d \log M / d t |_{\rm TD} \simeq 4 \times 10^{-10} 10^{\feoh_{\rm peak, TD} + 0.8} \hbox{ yr}^{-1}, 
\label{eq:sfr-tdp}
\end{equation}
   in the same way as above but with the metallicity, $\feoh_{\rm peak, TD}$ at which the iron-production by Type Ia SNe becomes important.  
   It is likely that the star forming region of Pop~IId is involved by such star bursts and/or affected by the yields produced by the latter CCSNe. 
   This is beyond the scope of the SAGA database and needs more data of stellar populations of higher metallicity ranges.  

\subsection{Transition between the lower- to higher-metallicity populations and change in the Initial Mass Function}\label{subsec:IMF}

Next we discuss the relationship between Pop~IIb and Pop~IIc.  
   They are observed to reside together in a wide metallicity range of $-2.6 \la \feoh \la -2.2$, with distinct spatial distributions.  
   Since the stellar evolution are basically determined by the mass and metallicity,  
   we ought to interpret their different abundance patterns in terms of a change of IMF, coupled with the dependence of SN yields on the mass of progenitors.   
   The discrepancies of mean enhancement are observed for Zn and Co, which are synthesized in the deepest shells during SN explosions as well as for C, which is formed during the quasi-static evolution and survived in the outermost part of core. 
   It is argued that the yields of larger $\abra{Zn,Co}{Fe}$ are ejected from hypernovae (HNe) with much greater explosion energies than normal CCSNe, which are expected for some of progenitors in upper mass range of CCSNe \citep{umeda05,tominaga07}. 
   For Pop~IIb, therefore, the IMF should be top heavy with a shallower slope in the mass range of CCSNe as compared with that for Pop~IIc.  



\begin{figure}
\begin{center}
\includegraphics[width=100mm,height=60mm,bb=50 50 410 295, clip]{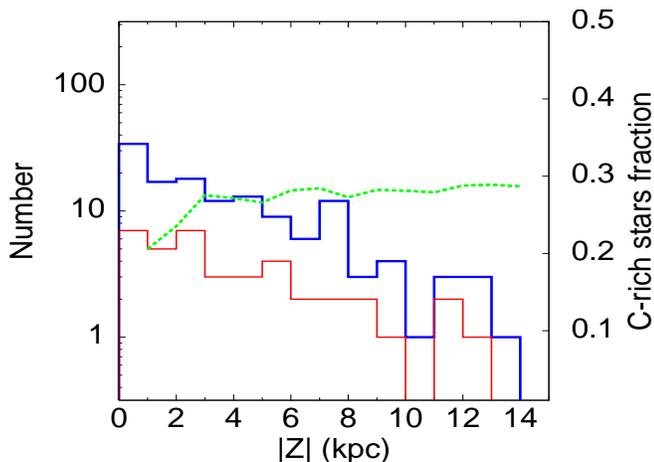}
\end{center}
\caption{
Spatial distributions along the vertical distance, $|Z|$, from the Galactic plane of carbon-enhanced ($\abra{C}{Fe} \ge 0.7$) giants (red histogram), giants with carbon abundance detected (blue histogram), 
in the metallicity range of $-3.1 \la \feoh <-2.2$.
   Also plotted are the fractions of the carbon-enhanced giants among the sample giants with the carbon abundances detected (green line) 
within the same vertical distance $|Z|$. 
}
\label{fig:cemp-distz}
\end{figure}

For stars in \EMP/\UMP\ divisions, the high-mass IMF with typical mass $\sim 10 \msun$ is derived from the statistics of CEMP stars, and also from the scarcity of low-mass survivors in comparison with the iron production by SNe \citep{komiya07,komiya09a}.  
   These conditions are related to the ratios between the low-mass stars and the intermediate-mass or massive stars.    
   Figure~\ref{fig:cemp-distz} compares the spatial distributions along the height, $|Z|$, from the Galactic plane for the giants with and without the carbon enhancement in the \EMP\ division ($\feoh > -3.1$).  
   The spectroscopic observations of these two groups of stars are usually performed separately, so that their relative numbers may be subject to the selection bias of observed targets.  
   And yet, the observed height distributions reflect the actual spatial distribution since the luminosity of giants varies greatly with the evolutionary stage and the distance can be known only after the determination of stellar parameters by the observations.  
   In this figure, the fraction of CEMP stars increases with the height to be nearly constant for $|Z| > 3$ kpc.   
   This is broadly consistent with the results by \citet{carollo12}, who derive the global trend of CEMP star fraction increasing with $|Z|$ for the sample stars of Sloan Digital Sky Survey.  
   The results indicates larger fraction of CEMP stars for Pop~IIb than for Pop~IIc at the same metallicity range, which is explicable in terms of the change of IMF from a high-mass to low-mass one \citep[][]{komiya07,suda13}.  

Accordingly, the IMF of Pop~IIb has to be both high-mass and top-heavy from large low-mass cut-off and large fraction of more massive CCSNe, while for Pop~IIc, the small fraction of CEMP stars and the contribution of Type Ia SNe indicate a low-mass IMF.   

\subsection{Transition between two lower-metallicity populations and the dilution of SN yields }\label{subsec:IMF}

Between Pop~IIa and Pop~IIb, the difference of mean enhancement is found only for Zn and Co, which is attributable to the difference in the contribution of HNe.  
   For Pop~IIa, lower metallicity demands greater dilution of the iron yield with the ambient gas than for Pop~IIb before it is incorporated into succeeding star formation.  

If the first SN is HN in a halo of sufficiently small mass, on the other hand, it may occur that the gas of host halo is blown off with the yield because of much greater explosion energies and of the lack of metals as coolant in the ambient gas \citep[e.g.,][]{machida05}.    
   Then, the yield ejected from the host halo undergoes greater dilution than in the host halo by the inter-halo gas, and then, can be incorporated again into more massive halos that later collapse. 
   In this case, the stars that form before the first SNe in the halos pre-polluted with the blown-off yield should be of lower metallicity than the stars that form with the yields from the first SNe diluted within the host halos.  
   We may conceive the origin of Pop~IIa with the feedback of HNe in their host halos.  
   In this connection, we note that the total sample stars in \UMP\ division remain about a half of those in \EMP\ division between $-3.3 < \feoh < -3.0$, which is too small to regard Pop~IIa as the precedent of all Pop~IIb.

\subsection{Two distinct modes of star formation and the sites} \label{subsec:two_modes}

The present results demonstrate two globally different modes of star formation during the early stages of the Galactic formation, which are distinguished in the IMF and the spatial distribution of low-mass stars surviving to data as well as in the metallicity. 
   One is for the lower-metallicity populations, IIa and IIb, which is characterized by high-mass and top-heavy IMF(s), and leaves behind their low-mass survivors, distributed all over the Galactic halo.  
   The other is for the higher-metallicity populations, IIc and IId, which is characterized by low-mass IMF(s) and brings the low-mass survivors in a flattened distribution toward the Galactic plane.  
   Although the change in IMF is usually discussed in relation to the difference in the metallicity, it is no longer applied in this case since the both modes are observed to work in parallel under the same metallicity.  
   It is natural to consider the IMF change in relation to the structural variations during the Galaxy formation which may give rise to the distinct distributions of low-mass survivors. 

In the framework of hierarchical structure formation of CDM scenario, we may conceive two different circumstances of star formation through the merger of halos.  
   One is in the mini-halos and halos of smaller mass ends, where gas is collapsed with the dark matter and condensed in their central regions.  
   The other is in pre-Galactic halo that grows sufficiently massive, in which gas, accreted and/or merged from other less massive halos, are condensed while moving in its larger gravity of the host halo. 
   These two sites are expected to bring about the star formation under different conditions. 
   In the former case, the star formation occurs under the self-gravity of gas, little influenced by the gravity of host halos. 
   In the massive pre-Galactic halo, on the other hand, gas accumulated through accretion and merging is subject to the shear of differential rotation and/or the tidal distortions of the gravity of the host halo, so that gas clouds tend to be elongated or flattened to have disk-like structures.  
   Accordingly, the fragmentation of gas clouds is postponed until a higher density is reached than in the halos of smaller masses, leading to low-mass IMF.  

The following picture of star formation history in the early Universe emerges from our study.  
   Pop~IIa and Pop~IIb stars are formed with a high-mass/top-heavy IMF in the host halos of relatively small masses, including the mini-halos first collapsed to birth stars; 
   the low-mass stars, currently observed, are after birth accreted onto the Galaxy at the merger of their host halos.  
   In this case, the low-mass stars are born mostly as the low-mass members of binary systems, and hence, their number is smaller in comparison with the iron produced by SNe.  
   On the other hand, Pop~IIc and Pop~IId stars are formed with a low-mass IMF out of gas, accumulated and flattened by the gravity of pre-Galactic halo that has grown sufficiently strong through merger. 
   The low-mass stars are formed in the flattened and/or disk-like orbits, although they may suffer from disturbances caused by the variations of rotation axis of pre-Galactic halo through the subsequent merging event(s).  
   In this connection, it is worth noting the different trends in the velocity dispersions of halo stars around $\feoh \simeq - 1.5$;   
   they remain essentially unchanged for lower metallicity, while monotonically decrease for higher metallicity with the border metallicity corresponding to the onset of formation of thick disk, as pointed out by \citet[][see also Carollo et al. 2010]{chiba00}.

\section{Conclusions}\label{sec:conclusions}

We have studied the detailed enrichment histories of individual elements, imprinted on the surface abundances of Galactic halo stars using a large sample stars of $\feoh \le -1$, collected by the SAGA database \citep{suda08,suda11}.  
   We have explored the metallicity dependence and spatial variation of the mean enhancements relative to iron with the aid of \WT\ for nine elements with the sufficient data, carbon, $\alpha$-elements (Mg, Si, Ca, and Ti), the odd-atomic number element (Sc), and the iron-peak elements (Co, Ni, and Zn). 
   We also discuss O, V, Cr and Mn, which are hard to draw statistically significant results because of discrepancies among the abundance data, in Appendix \ref{app:unanalyzed}.  
\red{
   In the analysis, we give due attention to the selection biases and systematic errors possibly involved in the data assembled from different authors.  
   In particular, we focus on the differential variations between the adjacent metallicity ranges in order to minimize these effects. }  

Our main conclusions are summarized as follows.

\noindent
(1)  We find steep variations, or breaks, of mean enhancement at three metallicities, centered at $\feoh$=$-1.8$, $-2.2$ and $-3.3$, among the sample stars;   
   across the most metal-rich break, all the elements except Co and Sc devoid of the abundance data for $\feoh >-2$ exhibit downturns in the mean enhancement, $\mabra{X}{Fe}$, with the metallicity by the same extent, $\Delta \mabra{X}{Fe} \simeq 0.08$ dex, within errors, while two other breaks are found for limited, but multiply, elements of C, Co, Zn and possibly Sc, and of Co and Zn, respectively. 
   Outside the breaks, no statistically significant changes of $\mabra{X}{Fe}$ are detected.  

\noindent
(2)  We identify four stellar populations, Pop~IIa, Pop~IIb, Pop~IIc and Pop~IId, which constitute dominant components in each metallicity range, divided by these breaks in order of increasing metallicity. 
   They are characterized by the different abundance patterns, and in particular, we find one-to-one correspondence to the average enhancements of Zn and Co, though for the latter, the most metal-rich break has to be confirmed by future observations.  

\noindent 
(3)  There are also variations in the scatter of distributions of $\abra{X}{Fe}$ among elements and for populations.  
   The scatters little differ between Pop~IId and Pop~IIc despite the shift of average enhancement.  
   It is as small as typical errors of observations for Ca and all heavier elements while much larger for the the lighter elements, C, Mg, and Si, studied.  
   Between Pop~IIb and Pop~IIc, some correlation is discerned for the scatters to increase with the variation of mean enhancement.  

\noindent
(4)   The Galactic halo stars, observed to date, are dichotomized by the spatial distributions along the vertical distance, $Z$, from the Galactic plane.    
   Stars of two lower-metallicity Pops~ IIa and IIb spread out to $|Z| \simeq 15$ kpc or more, and stars of two higher-metallicity Pops~IIc and IId are flattened near the Galactic plane.  
    In addition, we reveal that Pop~IIb and Pop~IIc coexist in the metallicity range of $-2.6 \la \feoh \la -2.2$, segregated spatially. 

\noindent
(5)  The variations in the abundance patterns have different origins.   
   The transition from Pop~IIc to Pop~IId is attributable to the delayed iron production by Type Ia SNe in the same star forming site. 
   The significant contribution of Type Ia SNe implies a low-mass initial mass function (IMF), and also, places a constraint on the specific star formation rate, which is as small as that in the thin disk of the present day Galaxy.  
   The difference between Pop~IIb and Pop~IIc is explicable in terms of the change in IMF, coupled with the mass dependence of SN yields, but not by the metallicity since they coexist over the wide metallicity range.  
   For Pop~IIb and also Pop~IIa, the enhancement of Zn and Co, synthesized by HNe from massive progenitors, indicates high-mass and top-heavy IMF along with a large fraction of CEMP stars among them.  

\noindent
(6)  The present result reveals two different modes of star formation during the early evolutionary stages of Galactic halos;  
   one with a high mass and top heavy IMF, which leaves the low-mass survivors in an extended distribution, and the other with a low-mass IMF, which results in a flattened distributions of the low-mass survivors.  
   In the context of hierarchical structure formation of CDM scenario, we may conceive of their sites as two different circumstances of star formation during the merger processes of halos.  
   One is in the halos of smaller masses before the merging and/or disruption, where stars are formed of their own gas in the central regions. 
   The other in the pre-Galactic halo, formed through the mergers, where stars are formed out of gas, accumulated through accretion and merger, while orbiting in the potential field of the host halo that has grown sufficiently strong.
   The different behavior of gas and stars during the merging process may give rise to the differences in the spatial distribution of low-mass stars survived to date.   

\red{
In this paper, we have exploited the SAGA database and our study is restricted by the quality and size of assembled data.  
   In particular, more data are necessary not only for $\feoh < -3.3$ but also for $\feoh > -2$ and $|Z| \ga 3$ kpc to inquire into the characteristics of lower-metallicity populations.  
   It is desirable to advance and refine our results with unbiased and homogeneous data of larger sample stars, such as expected from upcoming surveys and follow-up spectroscopic observations by Gaia and TMT.  
}

\section*{Acknowledgments}

This work has been partially supported by Grant-in-Aid for Scientific Research (18104003,23224004,25400233) from the Japan Society of the Promotion of Science.


\appendix

\section{Outliers} 
\label{app:outliers}

We survey the outliers with the anomalous abundances beyond $3 \sigma$ from the means for the sample stars.  
   In our analyses, we also exclude blue metal-poor stars from \citet{preston00} since they are thought to suffer from mass transfer in binaries. 

For the analyses of Mg and other $\alpha$-elements, excluded are the following 10 stars, 9 stars with the anomalous abundances of Mg and a giant, HE1424-0241, very deficient in other $\alpha$-elements in addition to the carbon-enhanced stars, discussed in \S~\ref{subsec:carbon}:    
\hfill\break\noindent 
[1,2] BS16934-002 and HD 29574; the former is deficient in neutron-capture elements \citep{aoki05,aoki07,fulbright00}: 
\hfill\break\noindent 
[3-6] G251-24, G4-36, BD+80\_245 and G139-8, deficient in Mg between $-2.1 < \feoh < -1.7$ \citep{stephens02,ivans03,carreta02}: 
   G251-24 is deficient in all $\alpha$-elements \citep{ermakov02} and also in Ba \citep{stephens02}: 
   G4-36 and BD+80\_245 are deficient in $\alpha$-elements (Mg, Si, Ca) and neutron-capture elements (Sr, Y, and Ba), and the former exhibits the enhancement of iron-peak elements \citep[Cr, Mn, Co, Ni, and Zn,][]{james00,carney97,ivans03}:
   G139-8 is known also as Li-deficient star \citep{norris97} and deficient in both Ba and Sr \citep{elliott05}. 
\hfill\break\noindent 
[7] CS22169-035 with $\feoh = -3.04$, deficient in Mg, Ca, Sc, and Ti as well as in Co, Ni and Zn \citep{cayrel04}, which is a mixed star with $\abra{N}{Fe} = 1.02$ \citep{spite05}. 
\hfill\break\noindent 
[8] CS22873-139 with $\feoh = -3.40$, poor in $\alpha$-elements (Mg and Ca) and with very low abundance of Sr \citep{spite00}. 
\hfill\break\noindent 
[9]  CS22952-015 with $\feoh = -3.42$, deficient in $\alpha$-elements (Mg, Ca, Ti), iron-peak elements (Cr, Mn, Ni) and neutron capture elements (Sr, Ba) \citep{cohen07,honda04}, which is a mixed star with $\abra{N}{Fe} = 1.31$ \citep{spite05}. 
\hfill\break \noindent 
[10]  HE1424-0241 with $\feoh = -3.87$, very deficient in Si, Ca and Ti and neutron-capture elements while normal in Mg and rich in Mn and Co \citep{cohen08}.  

The similar deficiency in $\alpha$-elements is also found for stars in dwarf spheroidal (dSph) galaxies \citep[e.g.,][]{tolstoy09}. 
   The halo stars, G4-36 and BD+80\_245, are suspected to be born in dwarf galaxies and deposited in the halo after the birth \citep[e.g.,][]{ivans03,klochkova11,venn12}.
   A star, Car-612, in Carina dSph is shown to possess the low ratios of $\alpha$ and neutron capture elements to iron similar to the stars of Ivans et al. star \citep{venn12}.  
   This is suggestive of the same origin for G251-24 and G139-8.  

For Si, we leave out CS22958-083 with an extraordinary high $\abra{Si}{Fe} = 1.18$ \citep{preston06} in addition to the above outliers.   
   Among the latter, BD+80\_245 and HE1424-0241 show unusually low enhancement of $\abra{Si}{Fe} = -0.11$ \citep{ivans03} and $-1.0$ \citep{cohen07}, respectively.
   For Ca, BD-80\_328 is removed because of very low abundance, $\abra{Ca}{Fe}=0.13$ at $\feoh = -2.03$ and much larger difference of 0.22 dex between the iron abundances from Fe I and Fe II lines than the average of $\simeq 0.055$ dex \citep{gratton00}.  
   Among the above outliers, G251-24, G4-36, and BD+80\_245 exhibit very low abundances of Ca, while the rest show not so large deviations from the means.  

For Sc, we exclude three stars, G4-36, BD+80\_245, and CS30338-089, with $\abra{Sc}{Fe}=-0.76$, $-0.42$ and 1.26, respectively \citep{ivans03,aoki07}. 

For the iron peak elements, we exclude G4-36 and BD+80\_245 with unusually high and low enhancements, $\abra{Zn}{Fe}=1.00$ and -0.42, respectively \citep{ivans03}.  
   For Co, we exclude HE1424-0241 with $\abra{Co}{Fe} = 1.03$ \citep{cohen08} in addition.  
   For Ni, we exclude HE1434-1442 and HE2232-0603 with the highest and lowest enhancements of $\abra{Ni}{Fe} =0.76$ and $-0.65$ at $\feoh = -2.39$ and $-1.85$, respectively \citep{cohen06}, and HD5223 with the second smallest enhancement of $\abra{Ni}{Fe}=-0.47$ at $\feoh = -2.06$ \citep{goswami06}. 
   For Zn, we exclude dwarfs from \citet{bonifacio09} since their Zn I abundances of dwarfs with $\feoh < -2.5$ are systematically larger by $\sim 0.2 \mhyph 0.4$ dex compared to giants.  
   We also exclude the abundance data from \citet{bihain04}, mostly lie in the metallicity range of $\feoh >-2$, including two dwarfs, HD166913 and HD132475, with the enhancements larger by $\sim 0.45$ dex than the means at $\feoh \simeq -1.5$.  


\section{Enrichment histories I: the rest of elements studied in the text}
\label{app:analyzed}




\begin{figure}
\begin{center}
\includegraphics[width=100mm,scale=1,bb= 50 50 626 315,clip]{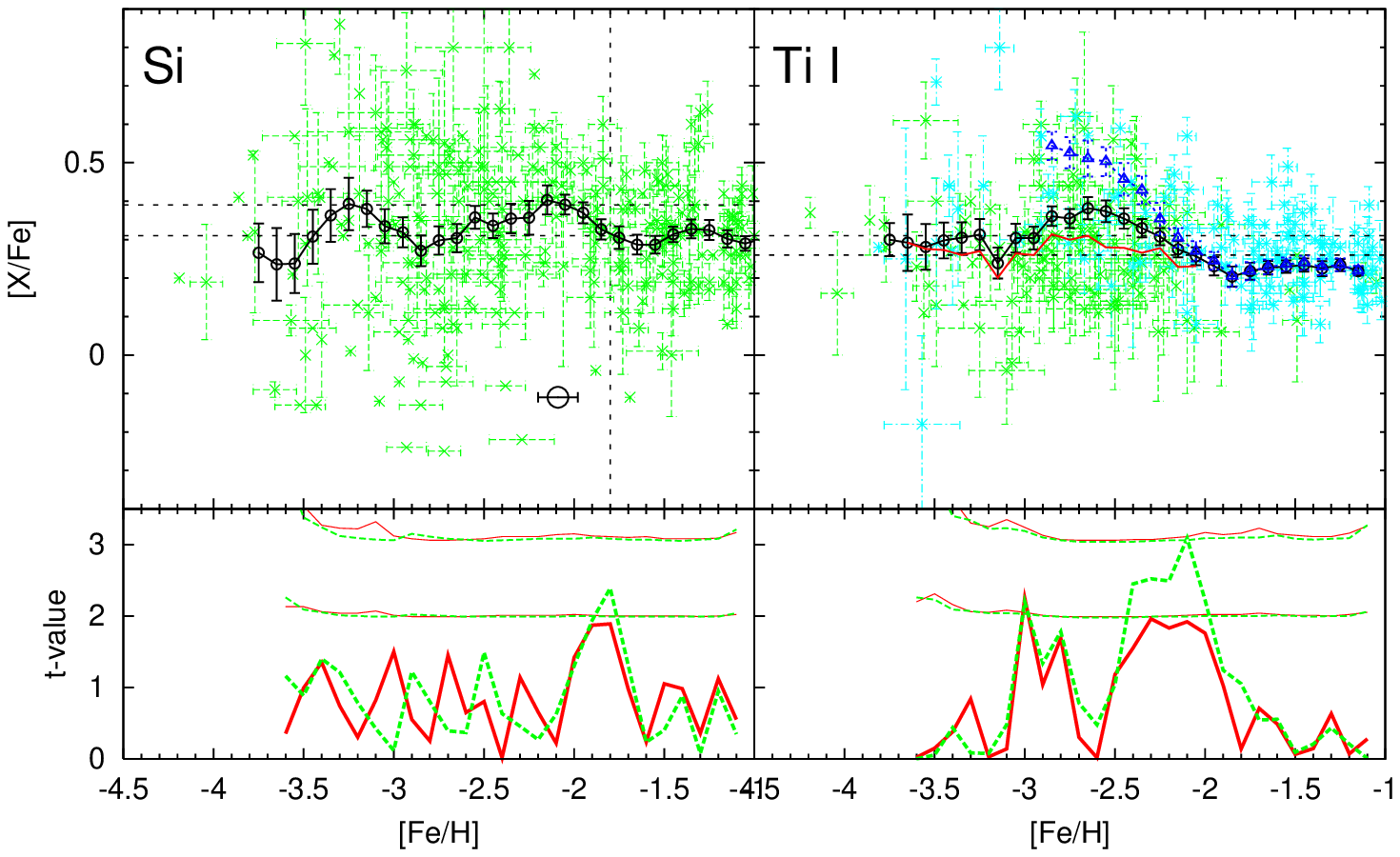}
\end{center}
\caption{ 
   The enrichment histories and the results of the \WT\ for Si and Ti I with the high-resolution abundances.  
   For Ti I, the enhancements of dwarfs and giants are denoted by blue and green symbols, and the mean enhancements for dwarfs and giants are also shown by triangle and red line, respectively.    
   Other symbols and lines have the same meanings as in Fig~\ref{fig:allfe}. }
   \label{fig:sitife} 
\end{figure}

\subsection{Silicon}

Figure~\ref{fig:sitife} (left panel) shows the enrichment history of Si for the sample stars with the high resolution abundances.  
   It displays the similar features to Mg and the downturn for higher metallicity across $\abra{Fe}{H} \simeq -1.8$ stands out by $2 \sigma$ level, although a flat part below it is degraded by undulations.  
   For the Si enhancement, however, decrease for higher effective temperature is reported among the metal-poor stars \citep{cayrel04,cohen04,lai08,for11};  
   \citet{preston06} argue to neglect the abundance data of stars with $T_{\rm eff} > 5800$ K.  
   
Fig.~\ref{fig:allfe} (top-right panel) shows the enrichment history without the stars of $T_{\rm eff} > 5800$ K.  
   Since the removed stars are concentrated in \EMP\ division, the depression disappears and the flat part extends down to $\feoh \simeq -2.4$. 
   Instead, an upturn appears below it, caused by a group of stars with $\abra{Si}{Fe} \ga 0.6$. 
   A bump around $\feoh \simeq -3.3$ is caused by the scarcity of the sample stars for $\abra{Si}{H} \le -2.6$ other than those from \citet{cayrel04} and \citet{lai08}.  
   The latter sample stars reach down to $\abra{Si}{H} \simeq -3.5$ and form a downturn for the lowest metallicity, though the sample size is not enough for the statistics.


\subsection{Calcium} 

The mean enhancement of Ca in Fig~\ref{fig:allfe} (middle-left panel) shares the same features as Mg and Si;  
   almost a flat distribution within $\sim 0.1$ dex with the feature, centered at $\feoh \simeq -1.8$.  
   The \STV\ gives the significance level of $2 \sigma$ with a wider peak than Mg and Si, probably related to significantly smaller scatter of enhancements.  
  There is a trend of decrease below $\feoh \la -3.3$ with low statistical significance in common to Si.

\subsection{Titanium}

For Ti, the discrepancy of abundances, measured by Ti I and Ti II lines, is reported \citep[e.g.,][]{johnson02,ishigaki10} as well as the dwarf/giant discrepancy.   
   It is argued that the Ti I abundances are subject to larger NLTE effects and the Ti II abundances are safely used \citep[e.g.,][]{bergemann11}. 
   
Fig.~\ref{fig:sitife} (right panel) illustrates the enhancement from the Ti I lines, where the mean enhancements are separately calculated and plotted for dwarfs and giants. 
   Dwarfs show a steep rise of $\mabra{Ti \;I}{Fe}$ below $\feoh \simeq -1.9$, which comes from the samples of $T_{\rm eff} \ga 6000$ K \citep{carreta02,cohen04,arnone05,preston06,lai08}.  
   In contrast, giants display $\mabra{Ti \;I}{Fe}$ rather small variations for $\feoh < -2$.  

Fig.~\ref{fig:allfe} (center panel) shows the enhancement from Ti II lines.  
   Here the downturn by $\sim 0.1$ dex is discernible across $\feoh \simeq -1.8$ in common with the other $\alpha$-elements, although the statistical significance is lower because of smaller sample size.  
   We also plot the mean enhancements of giants and dwarfs calculated separately for $\feoh < -2.3$, where the dwarf/giant discrepancy is apparent \citep{bonifacio09}.   
   Nearly flat distribution extends in either side of this feature, down to $\feoh \simeq -2.7$ for giants as well as up to $\feoh \simeq -1.5$.  
   In particular, there is no indication of the feature, centered at $\feoh \simeq -2.2$, again in accordance with other $\alpha$-elements.

\subsection{Scandium} 

The scandium is one of the odd atomic-number elements that lie in the valley of the smallest abundances between the $\alpha$- and iron-peak elements, together with vanadium, discussed in Appendix~\ref{app:unanalyzed}. 

The Sc abundances are derived from Sc II lines, which lie mostly in $\feoh < -2$ (only 15 stars in the range of $-2 \le \feoh < -1$).  
   In Fig.~\ref{fig:allfe} (middle-right panel), we see a decline of mean enhancement toward higher metallicity between $-2.2 \la \feoh \la -2.1$ by $\dave{Sc}{Fe} = 0.07 \pm 0.05$.   
   The \STV\ also peaks there for the both bin widths, but barely misses the threshold of $t_{2 \sigma}$ because of small sample number.  
   Outside this break, the mean enhancement remains nearly flat, although the data are too sparse for higher metallicity. 

The abundance data are taken mostly from the works, biased toward EMP stars \citep{cayrel04,cohen04,cohen06,cohen08,aoki05,aoki07,honda04,preston06,lai08}.  
   Among them, significant offsets exist from $\mabra{Sc \; II}{Fe} = 0.10 \pm 0.02$ \citep{cayrel04} to $0.27 \pm 0.05$ \citep{cohen04,cohen06,cohen08}.  
   On the other hand, \citet{johnson02} observe stars between $\feoh \simeq -3.0 \mhyph -1.5$, to give a difference of $\dmean{Sc}{Fe} = 0.03 \pm 0.05$ between \EMP\ and \VMP\ divisions, although the sample size is small.  

Accordingly, we only tentatively identify the downturn at $\feoh \simeq -2.2$ for Sc. 
   If confirmed, this distinguishes between the behaviors of odd and even atomic-number elements, Sc and Ti.   

\subsection{Cobalt}\label{appsubsub:Co}

The mean enrichment in Figure~\ref{fig:allfe} (left-corner panel) displays an upward trend toward lower metallicity, as previously reported \citep{mcwilliam95,ryan96,primas00,johnson02,cayrel04,takeda05,saito09,bergemann10}.  
   The mean enrichment increases from $\mabra{Co}{Fe} \simeq 0.1$ to 0.5 between $\feoh \simeq -2$ and $-3.5$, similarly to Zn, but the increase is not steady and nearly flat distributions develop between $-3.1 \la \feoh \la -2.4$ and for $\feoh \la -3.2$ within the errors. 
   The \WT\ also reveals the variations of $\mabra{Co}{Fe}$ across two metallicities, centered at $\feoh = -2.2$ and $-3.3$ by the statistical significance of $2 \sigma$ level. 
   Between and below these metallicities, $\mabra{Co}{Fe}$ may be regarded as constant within errors.  
   For $\feoh \ga -2$, the current sample size (5 stars) is too small to ascertain any features.

For Co, the sample stars are mainly from the works in favor of EMP stars \citep{honda04,cayrel04,cohen04,aoki05,cohen06,cohen08,lai08,bonifacio09}, while half of the data in \MP/\VMP\ divisions are from \citet{johnson02} which covers $\feoh \simeq -3 \mhyph -1.5$. 
   There are considerable differences in $\avabra{Co}{Fe}$ of \EMP\ division among these authors, ranging between $\avabra{Co}{Fe} = 0.14 \pm 0.05$ \citep{johnson02} and $0.29 \pm 0.03$ \citep{lai08} for giants. 
   For dwarfs, the averages are sill larger, ranging between $0.39 \pm 0.03$ \citep{bonifacio09} and $0.50 \pm 0.05$ \citep{cohen04}, which constitutes the dwarf/giant discrepancy \citep{bonifacio09}.  
   For \UMP\ division, the average ranges from $0.40 \pm 0.05$ \citep{cayrel04}, 
to $0.53 \pm 0.10$ \citep{bonifacio09}, which are also suspected of the dwarf/giant discrepancy.  

As for the downturn across $\feoh \simeq -2.2$, the data from \citet{johnson02} and \citet{preston06} give $\dave{Co}{Fe} = 0.02 \pm 0.12$ and $0.11 \pm 0.05$ dex, respectively,  
   which are consistent with the value in Table~\ref{table:average} within errors. 
   This suggests that the break is not an artifact of systematic errors.  


\subsection{Nickel}\label{appsubsub:Ni}

For Ni, \citet{cayrel04} and \citet{jonsell05} claim the mean enhancement close to $\mabra{Ni}{Fe} = 0$ for $\feoh \la -2$ and $\feoh \ga -2$, respectively.  
   On the other hand, \citet{ishigaki10} point out that $\abra{Ni}{Fe}$ in $-3 < \feoh < -2$ is slightly higher on average than for $\feoh \ga -2$. 
   \citet{nissen10} find a lower value of $\abra{Ni}{Fe}$ in the stars of lower $\abra{\alpha}{Fe}$. 

In Fig.~\ref{fig:allfe} (bottom-middle panel), the mean enhancement remains nearly flat within $\sim 0.1$ dex for all the metallicity, and the \WT\ detects a break at $\feoh = - 2.0$ - $-1.8$ by more than the 2$ \sigma$ level.  
   The trend agrees with $\alpha$-elements.  
   The enhancement of Ni, however, grows subsolar in \MP\ division in contrast to $\alpha$-elements, as pointed out by \citet{stephens02}.  
   In addition, there is an upturn for $\feoh \simeq -1.4$, for which the \STV\ reaches close to 2 $\sigma$ level. 
    Such a trend is also discernible for some of $\alpha$-elements.  

The sample stars above $\feoh \ga -2.2$ mostly come from the works which are more or less evenly distributed in these metallicity range \citep{fulbright00,johnson02,stephens02,gratton03}.  
   Thus, it is unlikely that the selection bias and/or systematic errors among the data from the different authors are the cause of the feature of $\feoh \simeq -1.8$.  
   On the other hand, the sample stars of $\feoh \la -2.5$ are mainly from the works biased for EMP stars \citep{cayrel04,cohen04,cohen06,cohen08,aoki05,honda04,lai08}. 
   Nevertheless, we hardly see any statistically significant variation around the joint metallicity of $\feoh \simeq -2.5$ - $-2.2$.   
   Small $t$-values of $\sim 1$ or smaller implies that the systematic errors among the authors are, if present, as small as, or less than, the observed errors. 

\section{Enrichment histories II: other elements}
\label{app:unanalyzed}
\subsection{Oxygen}\label{appsub:O}



\begin{figure*}
\begin{center}
\includegraphics[width= 200mm,scale=1,bb= 50 50 1058 315,clip]{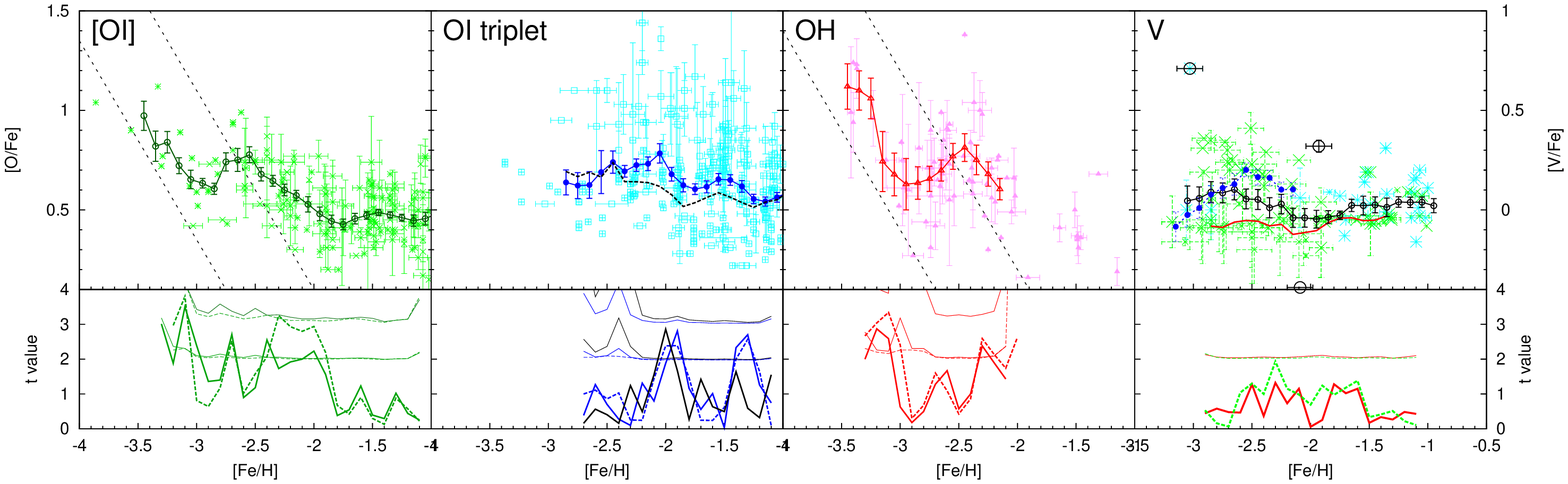}
\end{center}
\caption{The enrichment histories of oxygen and vanadium for the sample stars with the abundance data by the high resolution spectroscopy.  
   For oxygen, the abundances from [OI], OI triplet and OH lines are treated separately with the carbon enhanced stars excluded; 
   black lines in O I triplet panel denote the NLTE abundances with the data from \citet{fulbright03} removed. 
   For V, the enhancements, $\abra{V\; I}{Fe}$, of dwarfs and giants are plotted by light blue and green symbols; 
   red line denotes the mean enhancement for giants with $T_{\rm eff} < 5000$ K and blue symbols denotes the V II abundances. 
\label{fig:OV}}
\end{figure*}

For the oxygen abundance, three indicators of the forbidden [OI] lines, the OI triplet, and the OH lines are available. 
   Figure~\ref{fig:OV} shows the enhancements, measured by these three lines separately for the sample stars without the carbon-enhanced stars.  
   It is known that the LTE abundances from the different lines disagree for metal-poor stars \citep[e.g., Fig.~1 of][]{fulbright03}, and it is pointed out that the discrepancies reduce when the NLTE correction is applied \citep[e.g.,][]{gratton00,nissen02}.  
   Among them, the [OI] lines are argued as free from the NLTE effects and most reliable \citep[e.g.,][]{kiselman01,fulbright03}.  


For the [OI] abundances, the mean enhancement remains nearly constant at $\mabra{O}{Fe} \simeq 0.5$ for $\feoh \ga -2.0$.   
   Below this, it increases and forms a bump around $\abra{Fe}{H}\sim -2.5$.  
   This seems an artifact, arising from the selection bias by different authors, however; 
   the sample stars with $\abra{O}{H} \ga -1.8$ are crowded in the range of $-2.5 \la \feoh \la -2$, 
and rather sparse below $\abra{O}{H} \simeq -1.8$ (dotted line), both of which produce spurious upturn toward lower metallicity. 
   For $\feoh \la -2.7$, the mean enhancement decreases to develop a constant part with $\mabra{O}{Fe}$ larger by $\sim 0.15 \mhyph 0.2$ dex than the average in \MP\ division.  
   Another upturn for $\feoh \la -3$ is made also by the detection limit of $\abra{[OI]}{H} \ga -2.7$ \citep{spite05}. 

For the OI triplet, the enhancement exhibits different trend from the other two.  
   The mean enhancement remains nearly constant for $\feoh \ga -1.9$, larger by $0.1 \mhyph 0.2$ dex than for the [OI] lines.  
   For lower metallicity, it attains another nearly constant level after an upturn by $\sim 0.1$ dex.  
   The \WT\ shows breaks by more than $2\sigma$ level, centered at $\feoh \simeq -1.9 \mhyph -2.0$ and $-1.3$, respectively.  
   For the NLTE abundances (dashed line), the upturn by $\sim 0.1$ dex is discernible around $\feoh \simeq -2.0$, and below it, the mean enhancement is smaller by $\sim 0.1$ than the LTE abundance, near to $\mabra{[OI]}{Fe}$ in \MP\ division.   


For the OH lines, the mean enhancement traces the similar trend to the [OI] lines for $\feoh < -2$, including both the bump and the upturn due to the selection bias for $\abra{OH}{H} \ga -1.8$ in $\feoh \ga -2.5$ and to the detection limit of $\abra{OH}{H} \simeq -2.5$ for $\feoh < -3.0$ \citep{rich09,melendez01,lai08,israelian01}, respectively, and a flat part in between. 

Accordingly, we may infer the break between \MP\ and \VMP\ divisions and the absence of break of $\feoh \simeq -2.2$, in accordance with the other $\alpha$-elements, although a definitive conclusion waits for future observations with unbiased selections of target stars.  


\subsection{Vanadium}\label{appsub:V}

Fig.~\ref{fig:OV} (right panel) shows the enrichment histories both for the V I and V II abundances.  
   For V I, the mean enhancement (black line) displays an upturn for $\feoh <-2.2$, similar to Sc, with the \STV\ close to $t_{2 \sigma}$ at $\feoh = -2.3$ for 0.4 dex bin.  
    It is pointed out, however, that the V I abundance is sensitive to the effective temperature \citep{ivans99,fulbright00}, and a clear distinction shows up between giants and HB stars with $T_{\rm eff} \simeq 5300$ K as a border, reaching to $\sim 0.4$ dex for lower metallicity. 
   If restricted to the giants (red line), the mean enhancement decreases accordingly, and yet, the upturn is discernible along with nearly constant $\mabra{V \; I}{Fe}$ in the both sides.  
   In addition, there is an upturn for $\feoh \ga -1.7$ but with low statistical significance.  
   For V II, on the other hand, the enhancements exhibit quite different behavior, but only for 45 stars are available mostly in $\feoh \la -2$.  
   They are larger by $\sim 0.2$ dex or more than the V I enhancement of giants for $\feoh > -2.5$.  

Accordingly, it is hard to draw any statistically significant trend 
 although there is a vague sign of similar behavior with Sc.  

\subsection{Chromium and Manganese}\label{appsub:CrMn}

\begin{figure}
\begin{center}
\includegraphics[width=100mm,scale=1,bb=75 50 606 533,clip]{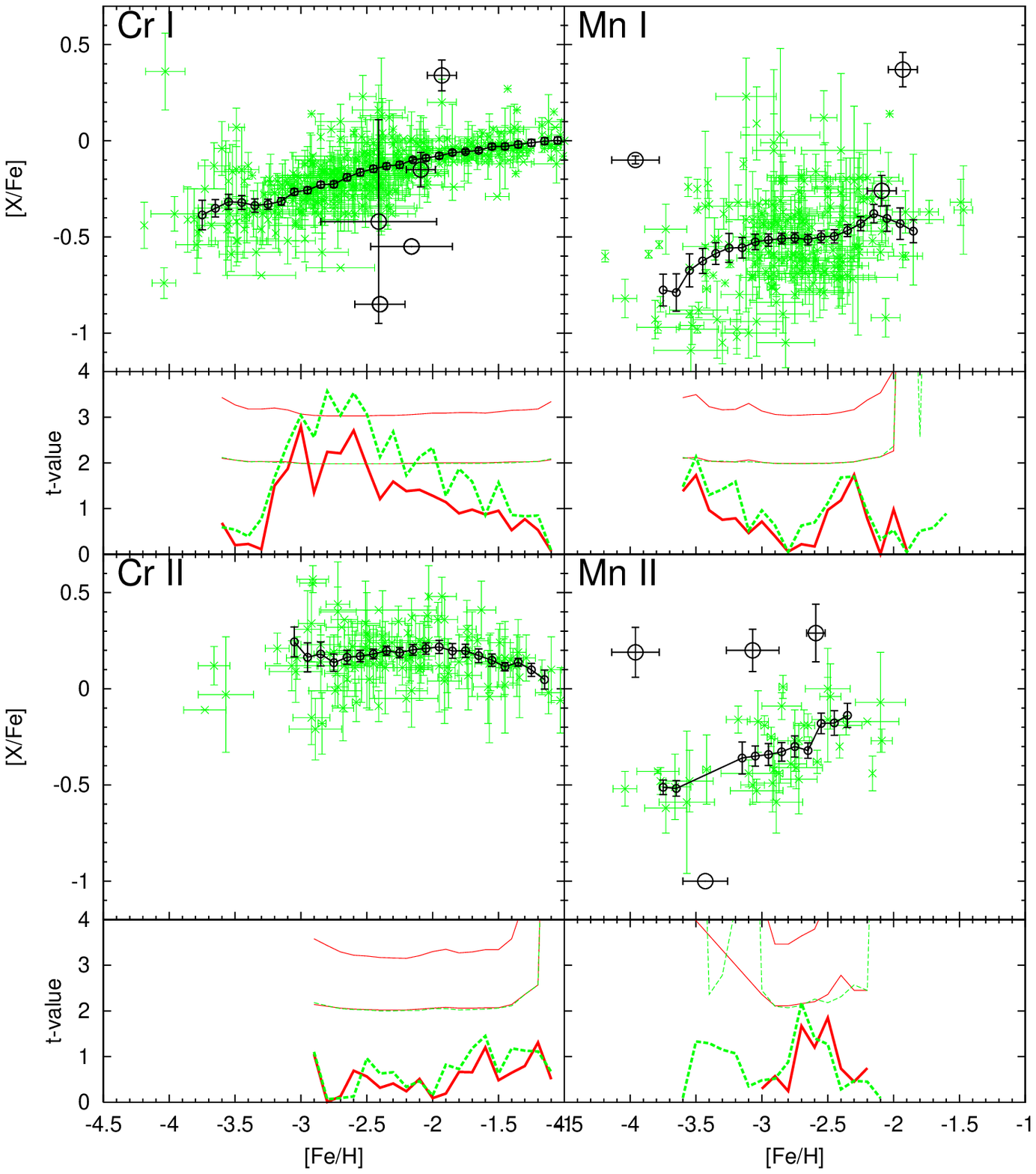}
\end{center}
\caption{The enrichment histories for Cr I and Cr II (left panels) and Mn I and Mn II (right panels) with the abundances by high dispersion spectroscopy.  
   Symbols and lines are the same as in Fig~\ref{fig:allfe}. 
  \label{fig:crIfe-t2}\label{fig:crIIfe-t2} \label{fig:crmnfe}}
\end{figure}


For the both elements, marked differences are reported between the abundances, derived from the lines of different ionizations \citep[e.g.,][]{johnson02,honda04,sobeck07,lai08,cohen08,bonifacio09, ishigaki10}. 
%
Figure~\ref{fig:crmnfe} compares the enrichment histories for the Cr I and Cr II abundances and for the Mn I and Mn II abundances, respectively. 

The enhancements from Cr I and Cr II lines display distinct behaviors;
   for Cr I, the mean enhancement increases with the metallicity more or less steadily over the whole range, while it remains nearly flat for Cr II. 
   The discrepancy between the Cr I and Cr II abundances enlarges from $\mabra{Cr \; I}{Cr \; II} \sim -0.1$ to $\sim -0.4$ as the metallicity decreases between $\feoh = -1$ and $-3$.  
   This is thought to point to the non-LTE effects \citep[e.g.,][]{sobeck07} and so the Cr II abundance to be more reliable \citep[e.g.,][]{bonifacio09}.  
    For Cr II, the mean enrichment remains nearly flat with a slight change in the slope around $\feoh \simeq -1.9$, although the \WT\ marks no significant differences partly because of large errors/scatters and of small sample size.  

For Mn, there is also a large discordance in Mn I and Mn II abundances, where $\abra{Mn \; I}{Mn \; II}$ enlarges for lower metallicity, similar to Cr. 
   Note that \citet{cayrel04} and some other authors add 0.4 dex to the Mn I abundance from the resonant triplet line, which makes the Mn I abundances suspect \citep{lai08}. 
   Different from Cr, however, the Mn II enhancements trace the trend of monotonic increase for higher metallicity, similarly to Mn I with an offset of 0.3 dex or more, although the data of Mn II abundances are too sparse to draw definite conclusions.  

The discrepancies of abundances from lines of different ionizations make statistically significant conclusions difficult, and yet, Cr II and Mn II provide an indication of different behaviors between even and odd atomic number nuclei, i.e., a nearly flat distribution for Cr enhancement common with $\alpha$-elements and Ni whereas a trend of enhancement increasing with the metallicity for Mn.    

{}

\end{document}